# Single Nanoparticle Dynamics in Opto-Thermal Tweezers: Resolving the Temporal Resolution of Depletion Force Trapping

Jinchao Chen,[1] Robert Talla Kontchou,[2] Saurabh Rai,[1] Guillaume Baffou,[1] Sylvain Blaize,[2] Quanbo Jiang,[2] and Jérôme Wenger[1,*]

[1] *Aix Marseille Univ, CNRS, Centrale Med, Institut Fresnel, AMUTech, 13013 Marseille, France*

[2] *Light, Nanomaterials, Nanotechnologies (L2n) Laboratory, CNRS EMR 7004, University of Technology of Troyes, 10004 Troyes, France*

** Corresponding author: jerome.wenger@fresnel.fr*

**Abstract:**

Optothermal tweezers enable the manipulation of a wide range of nano-objects through optically induced depletion forces. Despite significant advances, the temporal dynamics of optothermal trapping remain elusive, as existing methodologies rely almost exclusively on time and ensemble averaging. Consequently, stable trapping cannot be distinguished from local transient accumulation, where the time-averaged concentration increases but particles exhibit rapid, dynamic motion in and out of the trap. Here we investigate optothermal trapping with single-nanoparticle-level analysis and sub-millisecond temporal resolution. Our data resolve the elusive dynamics of 40 nm polystyrene nanoparticles trapped within depletion force potentials in polyethylene glycol solutions, enabling to differentiate the conditions leading to extended trapping times from those leading to transient localization. Numerical simulations corroborate our experimental findings, elucidating how the interplay between thermophoresis and diffusiophoresis governs nanoparticle dynamics. These insights deepen our mechanistic understanding of optothermal trapping and unlock opportunities for single-molecule studies, nanoscale assembly, and targeted drug delivery.

**Figure for Table of Contents**

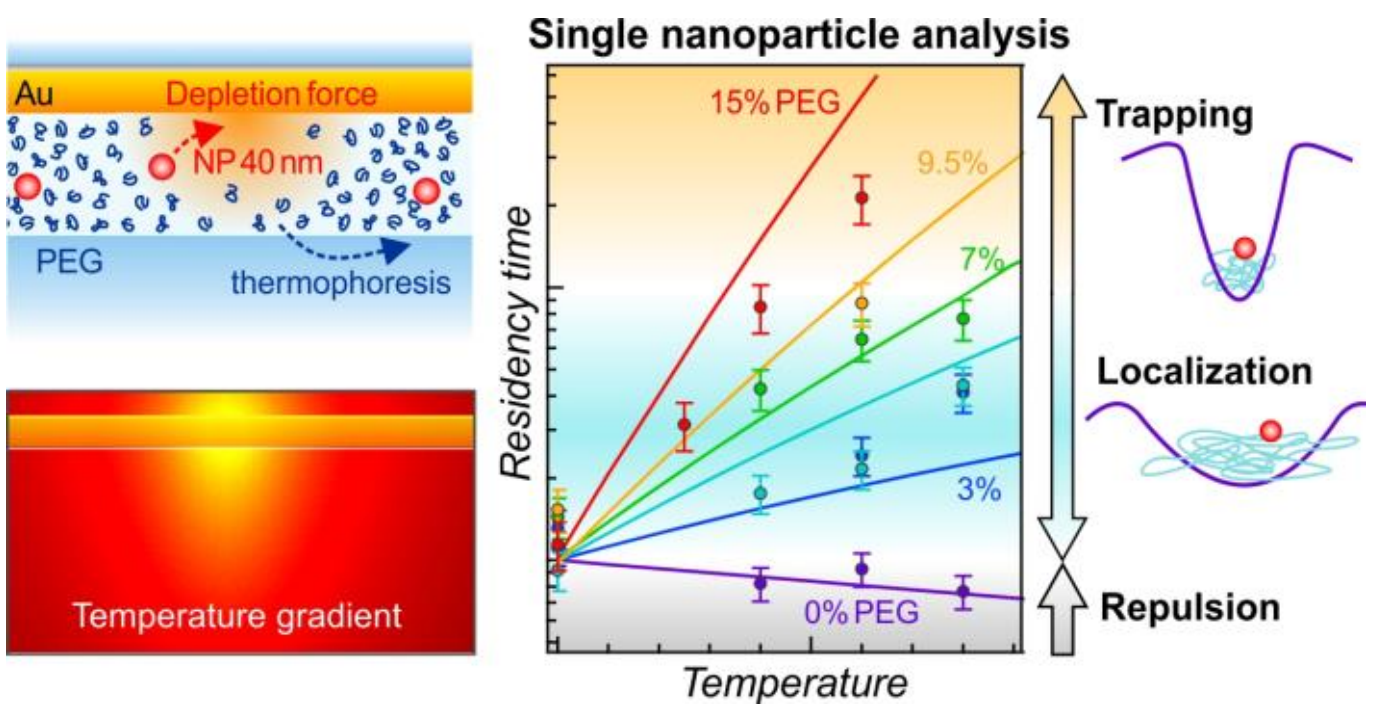

**Keywords:** nano-optical trapping, thermoplasmonics, thermophoresis, diffusiophoresis, optothermal tweezers

## Introduction

The inherent small size of nano-objects sets fundamental challenges on their manipulation, owing to their rapid µm²/ms diffusion and weak optical polarizability. Trapping of sub-500 nm nanoparticles using conventional diffraction-limited optical tweezers thus remains highly challenging.[1,2] To overcome these limitations, hybrid nanotweezers approaches have emerged integrating plasmonics,[3–7] metamaterials,[8–10] electrophoresis,[11–16] optofluidics,[17–21] and thermophoresis[22–26]. While these strategies have expanded the range of trappable nano-objects, they often introduce specific constraints. Plasmonic nano-optical tweezers require nanometer-scale gaps between metal structures,[27–33] which poses significant fabrication challenges. Anti-Brownian electrokinetic (ABEL) traps offer an alternative,[34,35] yet their dependence on fast active feedback and the requirement of a sufficient net electrical charge on the nano-object restrict their practicality.

Among recent advances, optothermal tweezers have emerged as a powerful yet simple approach for manipulating a wide range of nano-objects using optically induced thermodynamic forces.[36–46] These systems offer streamlined setups with significantly lower power demands compared to conventional optical tweezers. The physical mechanisms underlying optothermal tweezers rely on the interplay between two key phenomena: thermophoresis, which describes the directed motion of nano-objects along a temperature gradient, and diffusiophoresis, a related process in which colloids migrate along a gradient of solute concentration.[36,39] In presence of crowding agents, the combined occurrence of thermophoresis and diffusiophoresis leads to the depletion effect, wherein the thermophoretic migration of crowding agents generates a local concentration gradient. This gradient, in turn, induces diffusiophoresis of the target nano-objects, facilitating their precise localization and trapping.[41–45,47,48]

Despite substantial progress in optothermal trapping, current implementations rely almost exclusively on steady-state imaging via microscope cameras accumulating over a large number of nanoparticles for an extended time.[40–46] Consequently, the temporal dynamics of the trapping process remain largely inaccessible, preventing the distinction between stable trapping, where nanoparticles are firmly confined within the hotspot, and local accumulation, in which the time-averaged nanoparticle concentration increases locally but particles continue to exhibit dynamic, transient behavior in and out of the optothermal region. This ambiguity underscores a critical gap in our understanding: the lack of high-resolution temporal dynamic insights into the interplay between thermophoresis, diffusiophoresis and Brownian motion at the single nanoparticle level.[49] The challenge is further amplified by the presence of thermo-osmotic hydrodynamic flows at the nanoscale,[45,46,50] which are frequently overlooked in the analysis. Addressing this gap requires methodologies with single nanoparticle resolution capable of capturing both spatial confinement and dynamic behavior, enabling clearer insights on the nanoparticle residual movements.

Here we investigate optothermal trapping at the single nanoparticle level with sub-millisecond temporal resolution, leveraging a time-resolved fluorescence microscope that overcomes the limitations of the conventional cameras in temporal and ensemble averaging. Fluorescence burst analysis supplemented by fluorescence correlation spectroscopy (FCS) unveils the temporal dynamics of single 40 nm polystyrene (PS) nanoparticles undergoing optothermal trapping in polyethylene glycol (PEG) solutions. In molecular biophysics, FCS is a well-established technique to measure local concentrations, diffusion coefficients and fluorescence photophysics.[51,52] This technique has been adapted for optical tweezers to determine the trap stiffness,[31,47,53] in conjunction with related approaches such as confocal dynamic light scattering [54–56] and power spectral density analysis.[29] Additionally, FCS can measure the local temperature by change of the solution viscosity.[57,58]

Our experimental data demonstrate that trapping single 40 nm nanoparticles with prolonged residency times demands PEG concentrations above 7% and IR intensities exceeding 6 mW/µm². At lower PEG concentrations or IR intensities, the nanoparticles retain highly dynamic motion within the PEG-depleted region, exhibiting only modest increases in residency times compared to the unperturbed reference. Crucially, our results demonstrate that a local increase in nanoparticle concentration (as previously observed in camera-based studies) does not inherently signify stable trapping or extended residency times, underscoring a key distinction between transient accumulation and true confinement. Numerical simulations corroborate these experimental observations, illustrating how trapping efficiency scales with both IR intensity and PEG concentration. Collectively, these findings provide a comprehensive picture for understanding how depletion forces facilitate thermally driven trapping and govern nanoparticle diffusion dynamics.

**Results**

Figure 1a illustrates our experimental setup. The sample comprises 40 nm deep-red fluorescent polystyrene nanoparticles diluted in a Tris buffer solution containing varying mass fractions of 10 kDa PEG. This solution is introduced into a thin microfluidic chamber formed between two glass coverslips, with one coverslip coated with a 100 nm-thick gold film to act as a localized heat source under IR illumination. Gold was chosen for its chemical stability, absence of surface oxide layer, and broad accessibility.[59] The chamber thickness is maintained at approximately 10 µm to suppress convective flows.[46,60] A 1064 nm infrared (IR) laser is focused onto the gold film, generating a temperature gradient as approximately 2% of the IR power is absorbed by the gold film (Fig. 1b).[59,61] The fluorescence temporal dynamics induced by the motion of 40 nm nanoparticles is monitored using a confocal fluorescence microscope, with a 635 nm laser excitation and a single-photon counting avalanche photodiode. We have shown in a previous work that the IR illumination does not excite nor modify the fluorescence emission from the nanoparticles.[47] Detailed information about the samples and instruments are given in the Methods section.

Compared to earlier camera-based investigations,[41–46] our time-resolved fluorescence approach delivers single-nanoparticle sensitivity, enabling the observation of individual nanoparticle residency times within the optothermal trap and revealing previously hidden temporal dynamics. A consequence of this point-measurement is that no image can be recorded, precluding direct tracking of nanoparticle positions. To overcome this, we employ established single-molecule fluorescence microscopy techniques to characterize the nanoparticle motion by fluorescence burst analysis and fluorescence correlation spectroscopy (FCS).[51,52]

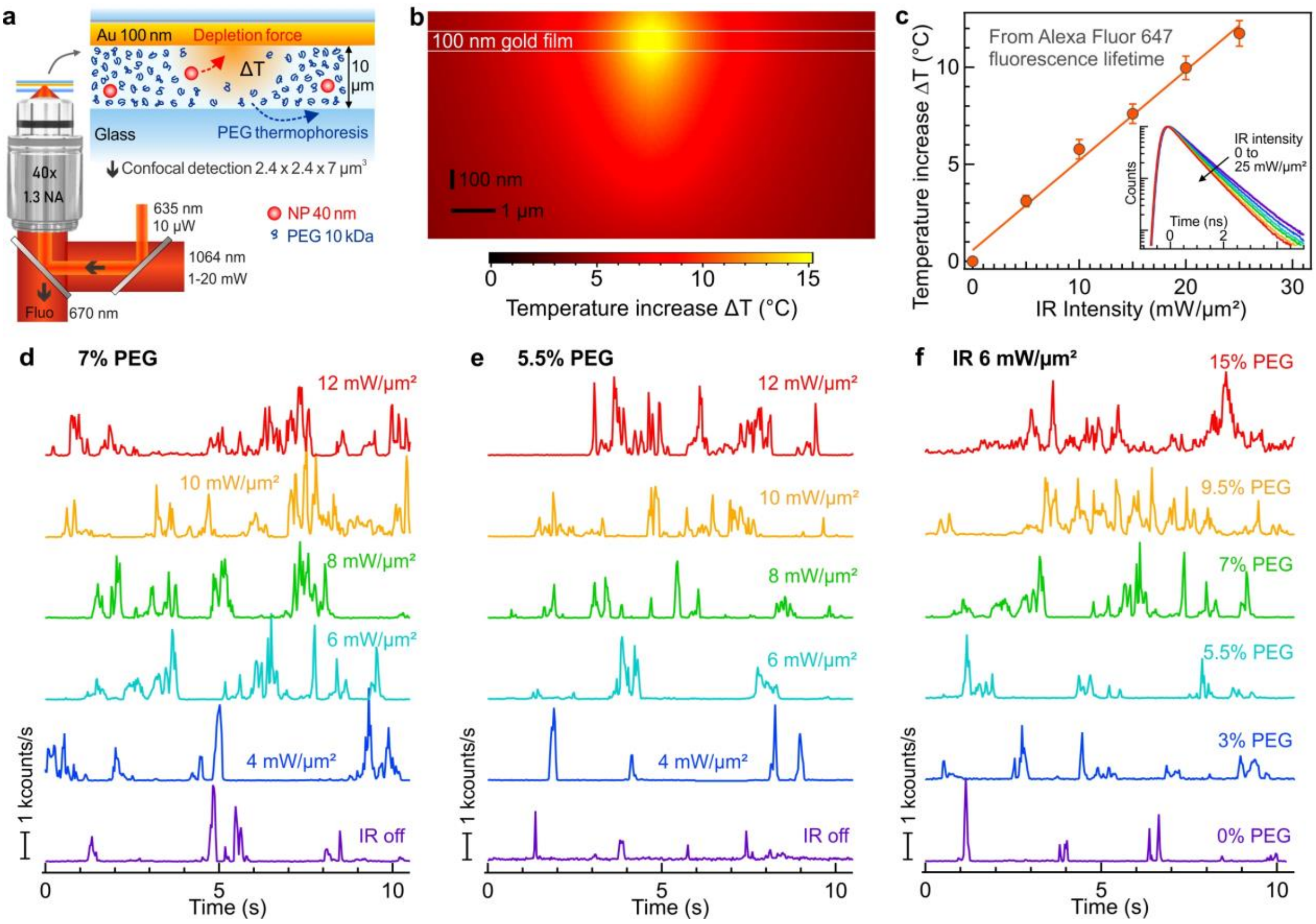

**Figure 1.** Opto-thermally driven depletion forces to manipulate single nanoparticles. (a) Scheme of the experimental configuration with an infrared beam for controlled local heating and a red laser beam to probe the fluorescence of the 40 nm polystyrene nanoparticles. Upon creation of a thermal gradient (orange center area), the PEG molecules in the solution undergo thermophoresis, which in turn creates depletion forces counterbalancing the thermophoresis of the polystyrene nanoparticles. (b) Numerical simulations of the temperature distribution along the XZ plane. The gold film is illuminated by a 10 mW laser beam with 550 nm $1/e^2$ waist at 1064 nm. (c) Experimental measurements of the local temperature increase averaged over the confocal detection volume as a function of the IR intensity,

using the fluorescence lifetime information of Alexa Fluor 647 dyes (insert) and the protocol derived in ref.[58,59] (d-f) Fluorescence intensity time traces for different IR intensities and PEG concentrations. The traces are shifted vertically to ease viewing. The sudden jumps indicate the presence of a single fluorescence nanoparticle. The binning time is 30 ms.

The optothermal trap operates through the synergistic interplay of thermophoresis and diffusiophoresis.[36] When the gold film is illuminated by a focused IR laser, a temperature gradient is established (Fig. 1b). Thermophoresis drives the thermophobic 10 kDa PEG molecules out of this temperature gradient, creating a local depletion in PEG concentration. Like the PEG molecules, the nanoparticles also undergo thermophoresis and tend to migrate toward colder regions. However, the PEG concentration gradient exerts an additional diffusiophoretic restoring force on the nanoparticles.[36,41–45] At sufficiently high PEG concentrations and temperature increases $\Delta T$, this depletion force can counterbalance the thermophoretic effect, enabling the local trapping of thermophobic nanoparticles at the hotspot. The theoretical framework underlying optothermal depletion trapping is detailed in Supporting Information Section 1 and will be further examined in Figure 4.

Our goal is to investigate the motion dynamics of single nanoparticles within the optothermal trap. We start by calibrating the local temperature increase $\Delta T$ as a function of the IR laser intensity (Fig. 1c). The temperature measurements are based on the fluorescence lifetime of Alexa Fluor 647 dyes, following the approach established in ref.[58,59] The temperature gain follows a linear dependence with the IR laser intensity consistent with the thermoplasmonic theory [62] and with prior observations.[59,63] We measure $\Delta T \sim 8$°C at 15 mW/µm² in presence of the gold film. A minor part of this temperature increase is due to the absorbance of the IR light by the aqueous buffer alone, an effect that is taken into account in the calibration in Fig. 1c. According to the study reported in the Supporting Information Fig. S2 of ref.[58], the temperature increase due solely to buffer absorption is 1.2°C at 15 mW/µm², a value consistent with an independent FCS-based study.[57]

Fluorescence time traces shown in Fig. 1d–f exhibit sudden intensity jumps corresponding to the motion of individual nanoparticles. In the absence of IR illumination, these bursts are short-lived, reflecting free nanoparticle diffusion. Upon increasing the IR intensity (Fig. 1d,e) or PEG concentration (Fig. 1f), longer events emerge, revealing optothermal trapping. This trapping also increases the frequency of single-nanoparticle bursts, which reflects a local concentration increase (Fig. 1d-f). Despite this phenomenon, the initial nanoparticle concentration in the picomolar range maintains our experiments in the single-nanoparticle regime even under elevated IR intensities and PEG concentrations, as validated by FCS characterization (Supporting Fig. S1). The case of higher initial nanoparticle concentrations leading to ensemble-averaging deviating from the single-nanoparticle regime will be examined later as we discuss the Supporting Fig. S10.

A simple first approach to quantify the nanoparticle temporal dynamics uses fluorescence burst analysis to determine the statistical distributions of burst durations (Fig. 2). Histograms for 7% mass fraction of PEG and its absence are shown in Fig. 2a,b. Other PEG fractions are presented in Fig. S2. In the absence of PEG, the histograms tend to get narrower with shorter bursts durations as the thermophobic nanoparticles undergo thermophoresis toward the colder regions. In presence of 7% PEG, longer bursts are visible, with a rising frequency upon IR intensity increase. To better visualize the data, we compute the complementary cumulative distribution functions (CCDFs) displayed in Fig. 2c,d and S3. The CCDFs represent the probability of observing a burst duration exceeding the time specified by the abscissa value. A comparison of Fig. 2c,d highlights striking differences between 7% PEG and its absence. Short burst durations are observed in absence of PEG across all IR intensities, showing the dominance of thermophoresis and the negligible contribution of optical gradient forces in our conditions. In presence of PEG and IR illumination, burst durations extend to several seconds, further increasing with higher PEG concentrations (Fig. 2e). This trend is quantitatively captured by the average burst duration (or equivalently median) shown in Fig. 2f. In the absence of PEG, the thermophoresis leads to a decrease of the burst duration as the IR intensity is increased. However in presence of PEG, the depletion force counteracts thermophoresis, restoring trapping stability and leading to longer burst durations with increasing IR intensity and PEG concentration. At the highest 15% PEG concentration, the data become more scattered, likely due to the high PEG density in the semi-dilute regime.[45]

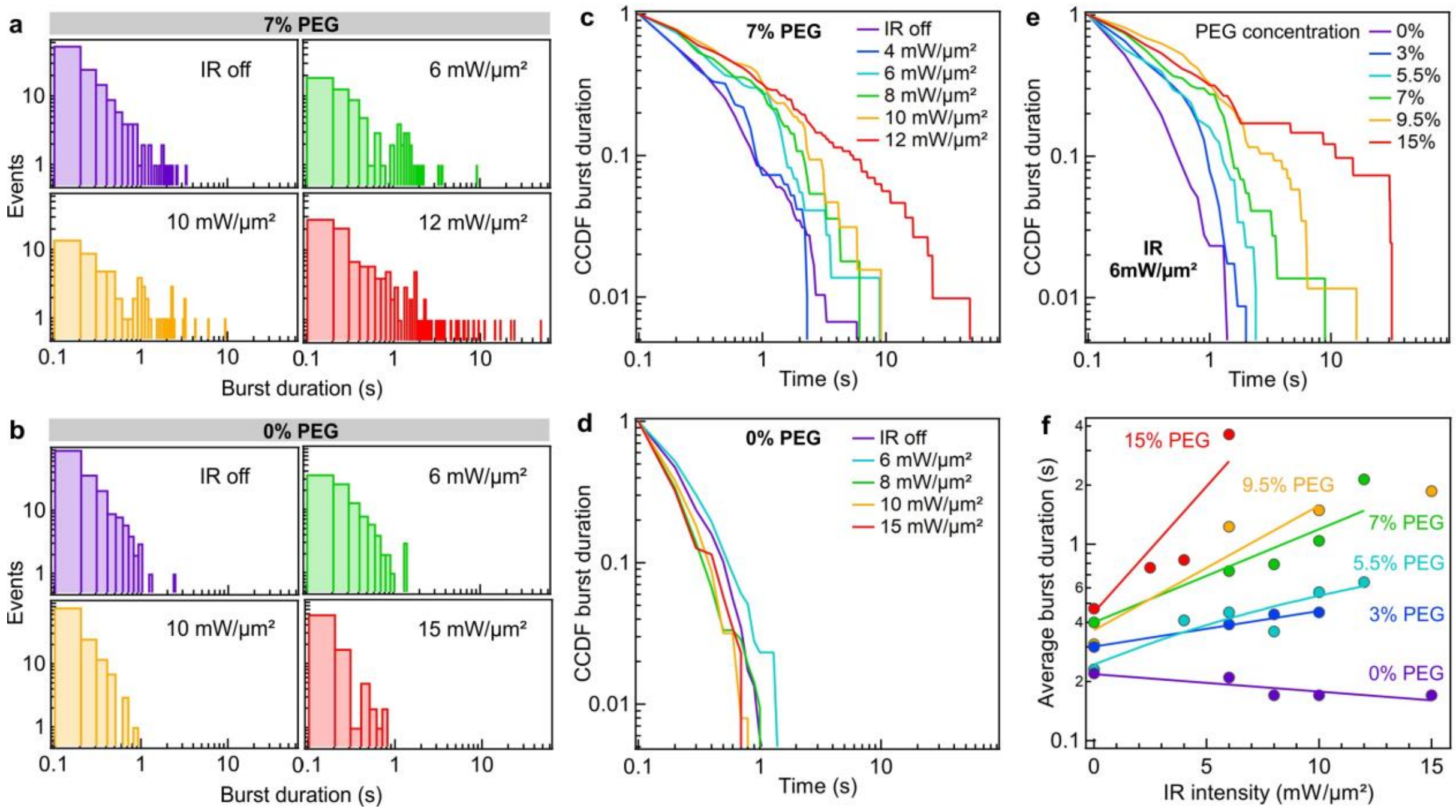

**Figure 2.** Fluorescence burst analysis reveals the residual motion dynamics of thermally-trapped nanoparticles. (a) Histogram of fluorescence burst duration recorded for 7% PEG concentration at different IR intensities. (b) Control histograms similar to (a) in absence of PEG. Supporting histograms are provided in Fig. S2. (c,d) Complementary cumulative distribution functions (CCDFs) computed from the data in (a,b). This corresponds to the probability of finding a burst duration longer than the abscissa value. Supporting CCDFs are provided in Fig. S3. (e) CCDF obtained for different PEG concentrations at a constant IR intensity of 6 mW/µm². (f) Average burst duration extracted from the histogram (a,b) plotted as function of the IR intensity for different PEG concentrations. The lines are numerical fits to the data.

The fluorescence burst analysis requires collecting a sufficient number of photons to distinguish nanoparticle signals from background noise, thereby limiting temporal resolution to the 10 ms range. Additionally, this method necessitates an intensity threshold, which may introduce bias towards the brightest bursts of longer duration. To overcome these limitations and achieve sub-millisecond temporal resolution, we implement FCS and compute the temporal autocorrelation of the fluorescence intensity traces.[51,52] The correlation data presented in Fig. 3a and S4 quantitatively assess the diffusion dynamics of the nanoparticles within the optothermal trap. Two distinct phenomena are visible on the FCS traces: a long 100-1000 ms component attributed to spatial diffusion in and out of the trap, and a short sub-ms component associated with interference fringes from the 635 nm excitation beam reflecting on the gold film (Supporting Information Fig. S5).[64] Since our primary focus is the dynamics within the optothermal trap, we concentrate on the long diffusion component, from which we extract the average FCS diffusion time $\tau_D$ displayed in Fig. 3b (see Materials & Methods for detailed FCS analysis).

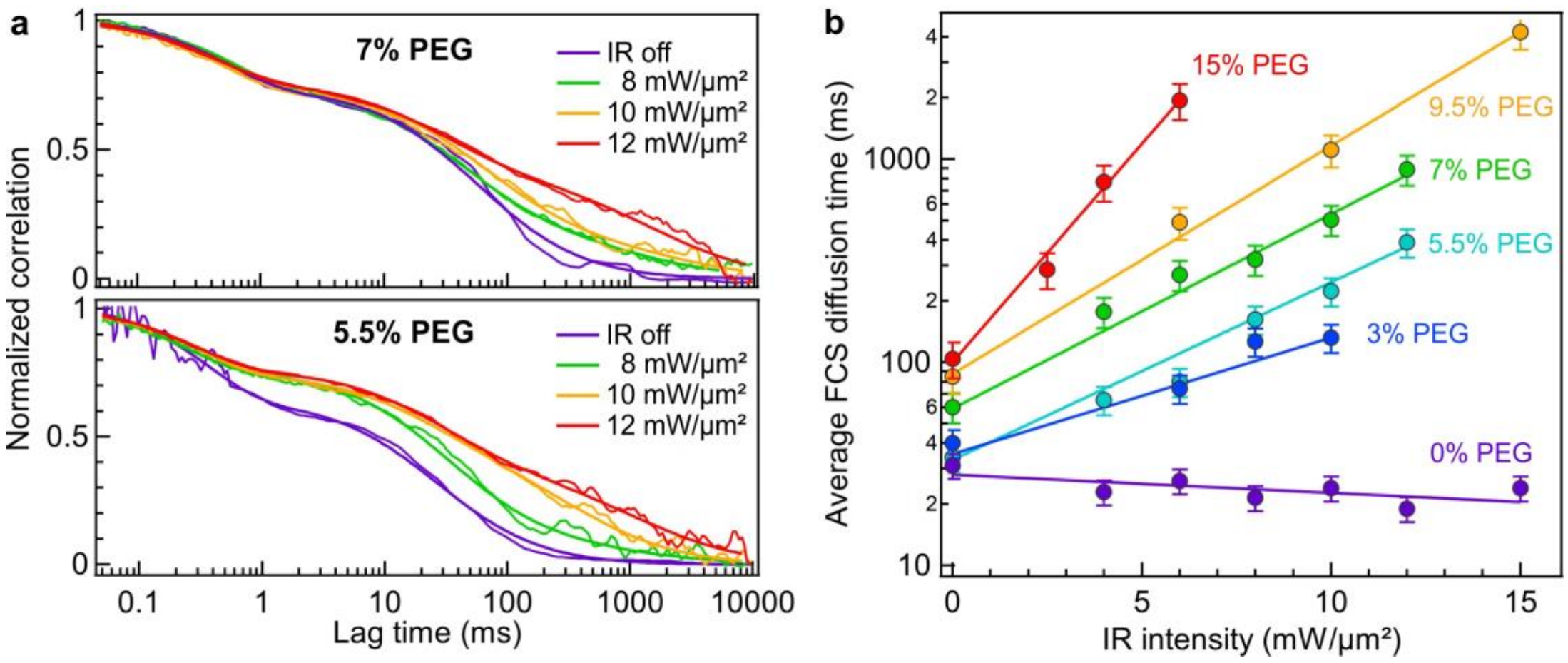

**Figure 3.** Fluorescence correlation spectroscopy data assess the nanoparticle motion dynamics within the optothermal trap. (a) Normalized FCS data (thin lines) and numerical fits (thick curves) corresponding to intensity traces in Fig. 1d,e for 7% and 5.5% PEG concentrations and different IR intensities. (b) Average diffusion time $\tau_D$ computed from the FCS fits as a function of the IR intensity for different PEG concentrations. The lines are numerical fits to the data. Since FCS involves no thresholding on the fluorescence intensity data (contrarily to the burst analysis), the values here differ from the average burst durations found in Fig. 2f, yet the qualitative conclusions remain identical.

In the absence of PEG, the nanoparticle diffusion time decreases with increasing IR intensity (Fig. 3b), confirming the thermophobic behavior of the nanoparticles as they migrate toward colder regions via thermophoresis. This observation also underscores the negligible role of optical gradient forces in our experiments, in consistency with the weak polarizability of 40 nm nanoparticles and the diffraction-limited focus of the IR beam.[65] In the absence of PEG and IR illumination, the 30 ± 2 ms FCS diffusion time is consistent with the 32 ms prediction based on Stokes-Einstein equation for a 40 nm nanoparticle size and a 1.2 µm waist of our confocal volume (Supporting Information Fig. S6).[51,52] This controls that the nanoparticles remain monodisperse in our experiments and do not aggregate.

In striking contrast, the PEG-mediated depletion force increases the FCS diffusion time with the IR intensity and the PEG concentration, improving the trap stability (Fig. 3b). As PEG also acts as a crowding agent, its addition alters the solution viscosity. We calibrate this effect by separate FCS experiments on CF640R fluorescent molecules in absence of IR illumination (Supporting Fig. S7), revealing that viscosity scales as $1 + 0.01{c_{PEG}}^2$, where $c_{PEG}$ is the mass fraction of PEG in %. At the highest 15% PEG concentration, viscosity increases by approximately 3.25-fold. When the IR laser is off and the temperature is uniform, the viscosity effect accounts well for the diffusion time increase (Supporting Fig. S8a). However, Figure S8b demonstrates that viscosity alone cannot explain the IR-dependent behavior observed in Fig. 3b. A major experimental result in Fig. 3b is that at PEG concentrations of 3–7% and IR intensities below 12 mW/µm², the FCS diffusion times remain below 1 second. This underscores the highly dynamic behavior of the nanoparticles rapidly entering and exiting the optothermal trap. Stable trapping with prolonged residency times emerges only at PEG concentrations above 7% and IR intensities exceeding 6 mW/µm².

To contextualize these observations, we compute the concentration gradients using a steady-state model described in the Supporting Information Section S1 and S2. Figure 4a displays the temperature profile $\Delta T(r)$ modeled as a Voigt function based on our calculations in Fig. 1b and the approach in ref.[44] The amplitude scales linearly with the IR intensity as calibrated in Fig. 1c. The PEG concentration follows the relation $c_{PEG}(r)/c_{PEG}(\infty) = \exp(-S_T^{PEG}\Delta T(r))$, where $S_T^{PEG} = 0.064$ K$^{-1}$ is the Soret coefficient for the 10kDa PEG molecules.[42,43,66] The positive value of $S_T^{PEG}$ reflects thermophobic behavior, resulting in a decrease of the PEG concentration in the hot region (Fig. 4b). Due to the PEG-mediated depletion forces, the effective Soret coefficient for the nanoparticles deviates from its intrinsic value $S_T^{NP}$= 0.075 K$^{-1}$, and is replaced by $S_{T,eff}^{NP} = S_T^{NP} - (S_T^{PEG} - 1/T)\, V c_{PEG}(\infty)$, [36,41–45] where $V = \pi\, \lambda^2\, d_{NP}$ with $d_{NP}$ is the nanoparticle diameter, and $\lambda$ the PEG depletion depth from the nanoparticle surface approximated by the 4.7 nm polymer gyration radius.[44,45,66] The additional diffusiophoresis term confers an effective thermophilic behavior on the nanoparticles, leading to their accumulation in the hot region (Fig. 4c).

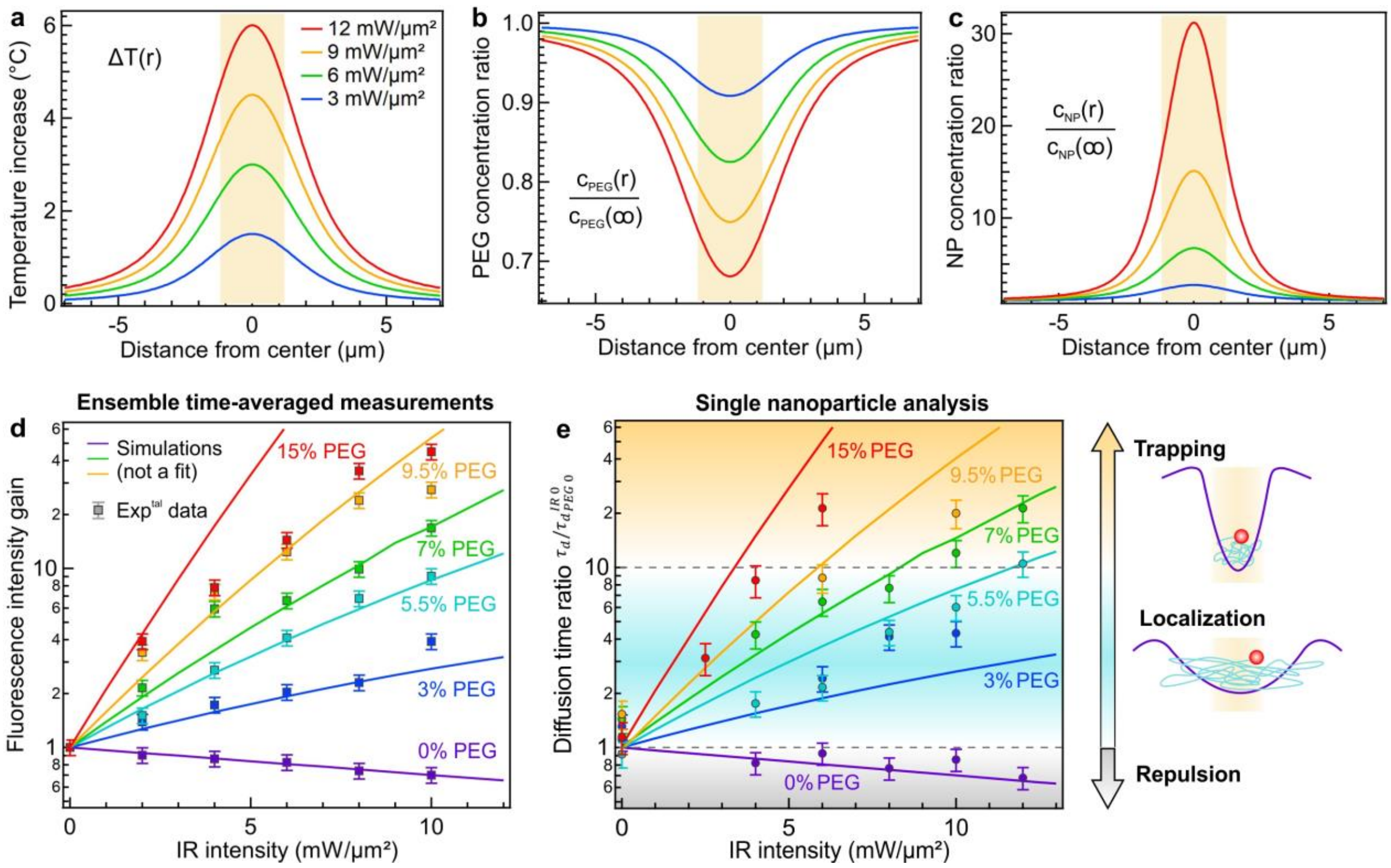

**Figure 4.** Numerical simulations support the experimental findings. (a) Temperature increase $\Delta T(r)$ as function of the distance to the laser spot center for different IR intensities. The shaded orange area in the center indicate the zone probed by the confocal microscope used in our experiments. (b) Relative PEG concentration profiles for the IR intensities used in (a) indicating that PEG molecule undergo thermophoresis to the cold areas ($S_T^{PEG}$= 0.064 $K^{-1}$).[66] (c) Relative PS nanoparticle concentration profiles for the IR intensities used in (a). Thanks to the depletion forces introduced by the local drop in PEG concentration, the nanoparticle thermophoresis ($S_T^{NP}$= 0.075 $K^{-1}$)[67] can be compensated, so the 40 nm PS nanoparticles accumulate in the center area. These calculations consider a 7% PEG concentration. (d) Fluorescence intensity gain averaged over the shaded central region in (c) as a function of the IR intensity for different PEG concentrations. The lines are numerical simulations, the markers are experimental data extracted from ensemble-averaged measurements at high NP concentrations (Fig. S10 and S11, 160x higher concentration than in Fig. 1-3). (e) Relative increase of the FCS diffusion time $\tau_d/\tau_{d\,PEG\,0}^{IR\,0}$ respective to the reference in absence of PEG and IR illumination. Lines are not a fit to the data, they represent the calculation results deduced from the simulations in (d) together with the calibrated dependence of the viscosity respective to the PEG local concentration (Fig. S7). The direct comparison of the non-normalized FCS diffusion time $\tau_d$ with the simulations is presented in Fig. S12. Different regions leading to time-averaged localization (low stiffness, short diffusion times) and trapping (higher stiffness, longer residency times) can be distinguished.

The steady-state predictions shown in the Supporting Fig. S9 indicate an enhanced nanoparticle localization with increasing IR intensity and PEG concentration, seen through an increase in the peak intensity and a reduction in the full width at half maximum. To experimentally verify these predictions, we perform experiments at 160x higher concentration and record the average fluorescence intensity from the photothermal trap (Supporting Information Fig. S10-S11). The results summarized in Fig. 4d align closely with the simulation predictions using as parameters the Soret coefficients for PEG and the nanoparticles, the nanoparticle diameter, the PEG gyration radius and the calibrated dependence of $\Delta T$ as function of the IR intensity. A deviation appears only for the highest 15% PEG concentration, as a possible consequence of the high PEG density in the semi-dilute regime,[45] in consistency with the observations in Fig. 2f. These experiments averaging over an ensemble of nanoparticles relate our single-nanoparticle observations in Fig. 1-3 to the conventional imaging approach.[41–45] They simultaneously validate the theoretical framework describing the interplay between thermophoresis and diffusiophoresis (Supporting Information section S1).[36]

The theoretical calculations of the nanoparticle concentration increase in Fig. 4c,d allow us to infer the relative increase in the trap residency time. In Fig. 4e, the computed lines consider that the relative increase in diffusion time $\tau_d/\tau_{d}{}_{PEG\,0}^{IR\,0}$ scales with the local concentration increase $c_{NP}/c_{NP}^{IR\,0}$ and the calibrated viscosity increase $\eta_{PEG}/\eta_{PEG\,0}$ (Fig. S6). The experimental data in Fig. 4e are normalized using the FCS diffusion times from Fig. 3b taken in absence of PEG and IR illumination. The non-normalized results are provided for comparison in the Supporting Information Fig. S12. Despite the simplicity of the model and the inherent challenges of the single-nanoparticle experiments, Figure 4e shows a remarkably good agreement between simulations and experiments across a wide range of IR intensities and PEG concentrations. Thermo-osmotic flows are not explicitly considered in our model, yet the agreement with experimental data (Fig. 4d,e) and earlier PEG studies [41,42] indicates their negligible contribution in our system.

As discussed in the Supporting Information section S3 for finite trap potential of small stiffness, the relative increase in diffusion time $\tau_d/\tau_{d}{}_{PEG\,0}^{IR\,0}$ scales linearly with the trapping potential depth $\Delta E/k_B T$.[68–70] Stable trapping generally demands $\Delta E/k_B T > 10$,[2,65] corresponding to residency times above 1 s and diffusion time ratios $\tau_d/\tau_{d}{}_{PEG\,0}^{IR\,0}$ above 10-fold (dashed line in Fig. 4e). Such conditions require PEG concentrations above 7% and IR intensities exceeding 6 mW/µm². In contrast, at lower PEG concentrations or IR intensities, the modest residency time increase reflects the dynamic motion of the nanoparticles rapidly entering and exiting the optothermal trap. We define the region where diffusion time ratios range from 1 to 10 as the “localization” regime. In this case, rapid movement of the nanoparticle still produces an apparent increase in the time-averaged local concentration (Fig. 4d), yet the conditions do not meet the requirements for stable trapping (Fig. 4e). Though the physics governing the optothermal trap remains the same, we believe that this distinction allows for a clarified understanding of the differences between transient localization over a broader region and stable trapping at the micrometer scale. Unfortunately, the rapid escape of nanoparticles from our optothermal trap precludes direct measurement of trap stiffness via FCS, which requires trapping durations exceeding 10 s. As a rough estimate, the stiffness of our optothermal trap is approximately one order of magnitude lower than that of plasmonic double nanohole tweezers. The characteristic trapping times and potentials observed in our system (Figs. 4e, S12, and S13) are comparable to those reported for self-induced back-action (SIBA) trapping in single nanoholes,[71] where trapping times of 100 ms to 3 s were observed for 100 nm PS nanoparticles. However, our optothermal trap requires approximately 10× higher IR powers, which we attribute to the use of a flat gold film rather than the focused-ion-beam-milled nanoholes employed in the SIBA study.

Crucially, comparing the data in Fig. 4d,e reveals that a local concentration increase (indicated by an elevated fluorescence intensity) does not inherently signify stable trapping and confinement at the sub-micrometer scale. Discriminating stable trapping from transient events requires careful analysis of temporal dynamics together with single-nanoparticle resolution. Additional experiments reported on Fig. S13 using 100 nm nanoparticles confirm these conclusions. The results for 10 kDa PEG extend to 4.6 kDa, as shown on the Supporting Information Fig. S14. The lower molecular weight of 4.6 kDa PEG results in a weaker response, which can be partially compensated by a higher concentration. However, these observations cannot be generalized to all crowding agents. For instance, we see that 40 kDa Dextran rather repels the PS nanoparticles from the hot area.

## Conclusions

In this work, we investigate optothermal trapping with unprecedented single-nanoparticle-level analysis and sub-millisecond temporal resolution. Our data uniquely resolve the elusive dynamics of nanoparticle motion within depletion force potentials. Stable optothermal trapping of single 40 nm polystyrene nanoparticles with residency times exceeding 1 second requires PEG concentrations above 7% and IR intensities exceeding 6 mW/µm², substantially higher than previously inferred thresholds. At lower PEG concentration and IR intensities, the nanoparticle localize around the trap center on a time-averaged basis, yet retain highly dynamic motion, rapidly diffusing in and out of the trapping region. Our findings underscore the need to differentiate between transient localization over a broader region and stable trapping, which ensures effective confinement at the sub-micrometer scale. This distinction is critical for advancing the field beyond ensemble-averaged camera-based approaches lacking temporal dynamics.

Numerical simulations corroborate our experimental observations, providing a cohesive framework for understanding how the interplay between thermophoresis and diffusiophoresis govern nanoparticle dynamics. These insights not only refine our mechanistic understanding of optothermal trapping but also open avenues for its application in single-molecule studies, nanoscale assembly, and drug delivery systems. Future work will explore the integration of optothermal tweezers with other manipulation modalities, such as electrophoretic or optofluidic controls, to expand the versatility and precision of nanoparticle trapping in complex environments.

## Materials and Methods

### *Sample preparation*

Deep red fluorescent carboxylate-modified polystyrene nanoparticles of 40 nm diameter are purchased from Thermofisher Invitrogen (ref F8789, absorption/emission maxima at 660/680 nm). For the trapping experiments, the fluorescent nanoparticles are diluted by 800,000x to a final concentration of $10^9$ particles/mL (about 1.6 pM) to ensure no more than one nanoparticle is present in the thermophoretic trap at a given time. The nanoparticles are sonicated for 20 min right before the experiments to avoid observing nanoparticle aggregates. The absence of aggregation is checked by fluorescence correlation spectroscopy (FCS) experiment, which confirm the hydrodynamic radius measured by FCS corresponds to the 40 nm diameter of the nanoparticles.

All dilutions are performed in ultrapure water (Merck Millipore DirectQ-3 UV, conductivity 18.2 MΩ.cm) with 50 mM Tris (Sigma Aldrich ref 252859) and HCl buffer at pH 7.7. Poly-ethylene glycol of 10 kDa molas mass (Sigma Aldrich ref 92897) is added to the solution with a given mass fraction. No salts are added to the solution. To further prevent adsorption to the glass and gold surfaces, 0.5% of Tween20 (Sigma Aldrich P1379) is added to the solution. Experiments exhibiting non-specific adsorption, identified by nanoparticle retention after IR beam interruption, were discarded.

The 100 nm-thick gold films used in this study are deposited on 150 µm thick borosilicate glass (Menzel Gläser #1) by electron-beam evaporation (Bühler Syrus Pro 710). A 5 nm chromium adhesion layer is used between the gold film and the glass substrate, yet this adhesion does not influence our results as we use a reverse illumination where the laser beam first interacts with the gold film (Fig. 1a). The absorbance of the gold film is about 2% at 1064 nm,[61] which is taken into account by the calibration in Fig. 1c. Optically thick layers of chromium and titanium absorb approximately 20 times more at this wavelength, experiments using these metals need to adjust the IR power accordingly.

For the trapping experiments, 1.5 µL of the PEG solution containing the fluorescent nanoparticles is sandwiched between a borosilicate glass coverslip and the gold film. This volume leads to a 10 µm height of the solution, which is chosen to avoid thermal convection and is verified by confocal microscopy. In our configuration, the optothermal trap extends over distances exceeding 100 nm, making Faxen's corrections unnecessary. To avoid evaporation, the sample is sealed and a 20 µL water droplet is added on top of the substrate holding the gold film to further reduce evaporation. Experiments could be performed over 60 min duration without noticing significant changes.

***Experimental setup***

The optical microscope is based on an inverted confocal microscope with a high NA microscope objective (Zeiss Plan-Neofluar 40x, NA 1.3, oil immersion). To generate the temperature gradient on the gold film, we use a continuous wave 1064 nm laser (Ventus 1064-2W) focused by the microscope objective to a 1 µm spot diameter. The infrared illumination intensity expressed in this paper (in mW/µm²) corresponds to the position of the gold film, taking into account the 50% transmission of our illumination optics at 1064 nm. To excite the fluorescence from the polystyrene nanoparticles, we use a 635 nm laser beam spatially overlapped with the IR beam (Picoquant LDH-P-635). To ensure an fluorescence illumination profile covering the spatial extension of the thermal trap, the beam diameter is reduced to 1.6 mm before the microscope objective, so that the 635 nm laser spot size features a $1/e^2$ waist of 1.2 µm on the sample plane (see a complete characterization in Fig. S6). The red laser intensity used in this work is 6 µW/µm². Experiments performed on immobilized nanoparticles confirm that photobleaching is negligible in our conditions (Supporting Information Fig. S15).

The fluorescence from the nanoparticles is collected by the same microscope objective in epi-configuration. A set of dichroic mirrors, long pass filters, 100 µm confocal pinhole and bandpass filters in the 650-750 nm spectral range rejects the scattered laser light and ensures that only the fluorescence light is detected. The detection is performed with a Perkin Elmer SPCM-AQR-13 single-photon counting avalanche photodiode. The photodiode output is connected to a time correlated single photon counting module (Picoquant Picoharp 300 with PHR 800 router) with time-tagged time-resolved (TTTR) option.

***Fluorescence data analysis***

All the fluorescence time traces are analyzed with the Symphotime 64 software (Picoquant) using the in-built functions of burst analysis and FCS. The FCS correlation data is interpolated using a 2-species 3-dimensional diffusion model with an additional term to account for the presence of the gold film acting as a mirror:[51,64]

$$G(t) = \frac{1}{N}\left(1 + Ae^{\left(\frac{-t}{\tau_f}\right)}\right)\left[\frac{\rho_1}{\rho_1+\rho_2}\left(1+\frac{t}{\tau_{D1}}\right)^{-1}\left(1+\frac{t}{\kappa^2\tau_{D1}}\right)^{-0.5} + \frac{\rho_2}{\rho_1+\rho_2}\left(1+\frac{t}{\tau_{D2}}\right)^{-1}\left(1+\frac{t}{\kappa^2\tau_{D2}}\right)^{-0.5}\right]$$

where $\mathrm{N}$ is the average number of nanoparticles in the observation volume, $\rho_1$ and $\rho_2$ are the respective weight for each species with diffusion times $\tau_{D1}$ and $\tau_{D2}$. The presence of the gold film reflects the laser light, leading to interference fringes as shown on Fig. S5. To account for the nanoparticle hoping between interference fringes, we add a $1 + A\,exp(-t/\tau_f)$ term following the approach in [64] where $\tau_f$ is the characteristic diffusion time between interference fringes. κ corresponds to the aspect ratio of the axial to the transversal dimension of the detection volume, which was kept constant to 5 following the established practice in FCS.[51,64] The FCS diffusion time considered in the main article corresponds to the amplitude-averaged value between $\tau_{D1}$ and $\tau_{D2}$ given by $\tau_D = (\rho_1\tau_{D1} + \rho_2\tau_{D2})/(\rho_1 + \rho_2)$. Similar conclusions could be obtained by considering only the longer component $\tau_{D2}$ or by using a 1-species model, yet we find that the 2-species approach generally yields better fits (see Supporting Information Fig. S16). The rationale behind the 2-species model could be related to differences in movement along the transverse and vertical directions.

***Numerical simulations of temperature distributions***

Electromagnetic and thermal simulations were performed using COMSOL Multiphysics with the Electromagnetic Waves, Beam Envelopes and Heat Transfer in Solids and Fluids modules. Convective effects were neglected due to the micrometer thickness of the liquid layer. Open boundary conditions were implemented using an infinite element domain, with fixed ambient temperature at the outer boundaries and continuity at material interfaces. Symmetry conditions were applied to reduce computational cost: one symmetry plane along the polarization direction and one antisymmetric plane normal to it for the electromagnetic problem, and two symmetry planes for the thermal model. The dielectric function of gold was taken from Johnson and Christy,[61] while all other material parameters were obtained from the COMSOL library. The absorbed power density was defined as $Q_{abs} \propto Im(\varepsilon)|E|^2$ and used as a volumetric heat source. A mapped mesh with extrusion was employed, with a minimum mesh size below 5 nm in the near-field region and gradual coarsening away from the structure. The system was excited at 1064 nm with a linearly polarized Gaussian beam (waist 550 nm). Although the structure is geometrically axisymmetric, the incident polarization breaks this symmetry at the electromagnetic level. However, owing to the lateral homogeneity of the environment, both the absorbed power distribution and the resulting temperature field remain effectively polarization-independent. The reference field $E_0$ is defined at the beam center. Mesh convergence and domain size independence were verified to ensure numerical accuracy.

**Supporting Information**

Theory of steady-state concentration gradients; Numerical computations of concentration gradients; Escape time from a trap potential; Average number of nanoparticles from FCS; Supplementary fluorescence burst duration histograms; Supplementary complementary cumulative distribution functions; Supplementary FCS data for different PEG concentrations; Interference fringes in the vicinity of the gold mirror; Calibration of the confocal detection volume; Viscosity increase in presence of PEG; Influence of the viscosity on the residence time; Simulations of concentration gain and localization width; Ensemble-averaged intensity data; Intensity data in absence of PEG; Simulations of FCS diffusion times; Experiments for 100 nm nanoparticles ; Extension to other configurations ; Control for negligible photobleaching; Control of FCS fit results.

**Acknowledgments**

This project has received funding from the Agence Nationale de la Recherche (project ANR-24-CE42-0846), the Graduate School NANO-PHOT (École Universitaire de Recherche, grant ANR-18-EURE-0013), and the Program QuanTEdu-France (grant ANR-22-CMAS-0001 France 2030).

**Conflict of Interest**

The authors declare no conflict of interest.

**Data Availability Statement**

The data that support the findings of this study data are available from the corresponding author upon request.

**References**

(1) Yang, M.; Shi, Y.; Song, Q.; Wei, Z.; Dun, X.; Wang, Z.; Wang, Z.; Qiu, C.-W.; Zhang, H.; Cheng, X. Optical Sorting: Past, Present and Future. *Light Sci. Appl.* **2025**, *14*, 103.
(2) Juan, M. L.; Righini, M.; Quidant, R. Plasmon Nano-Optical Tweezers. *Nat. Photonics* **2011**, *5*, 349–356.
(3) Ren, Y.; Chen, Q.; He, M.; Zhang, X.; Qi, H.; Yan, Y. Plasmonic Optical Tweezers for Particle Manipulation: Principles, Methods, and Applications. *ACS Nano* **2021**, *15*, 6105–6128.
(4) Zhang, Y.; Min, C.; Dou, X.; Wang, X.; Urbach, H. P.; Somekh, M. G.; Yuan, X. Plasmonic Tweezers: For Nanoscale Optical Trapping and Beyond. *Light Sci. Appl.* **2021**, *10*, 59.
(5) Gordon, R. Future Prospects for Biomolecular Trapping with Nanostructured Metals. *ACS Photonics* **2022**, *9*, 1127–1135.
(6) Crozier, K. B. Plasmonic Nanotweezers: What's Next? *ACS Photonics* **2024**, *11*, 321–333.
(7) Kotsifaki, D. G.; Chormaic, S. N. Plasmonic Optical Tweezers Based on Nanostructures: Fundamentals, Advances and Prospects. *Nanophotonics* **2019**, *8*, 1227–1245.
(8) Shi, Y.; Song, Q.; Toftul, I.; Zhu, T.; Yu, Y.; Zhu, W.; Tsai, D. P.; Kivshar, Y.; Liu, A. Q. Optical Manipulation with Metamaterial Structures. *Appl. Phys. Rev.* **2022**, *9*, 031303.
(9) Xu, Z.; Crozier, K. B. All-Dielectric Nanotweezers for Trapping and Observation of a Single Quantum Dot. *Opt. Express* **2019**, *27*, 4034–4045.

(10) Yang, S.; Hong, C.; Jiang, Y.; Ndukaife, J. C. Nanoparticle Trapping in a Quasi-BIC System. *ACS Photonics* **2021**, *8*, 1961–1971.
(11) Ndukaife, J. C.; Kildishev, A. V.; Nnanna, A. G. A.; Shalaev, V. M.; Wereley, S. T.; Boltasseva, A. Long-Range and Rapid Transport of Individual Nano-Objects by a Hybrid Electrothermoplasmonic Nanotweezer. *Nat. Nanotechnol.* **2016**, *11*, 53–59.
(12) Ndukaife, J. C.; Xuan, Y.; Nnanna, A. G. A.; Kildishev, A. V.; Shalaev, V. M.; Wereley, S. T.; Boltasseva, A. High-Resolution Large-Ensemble Nanoparticle Trapping with Multifunctional Thermoplasmonic Nanohole Metasurface. *ACS Nano* **2018**, *12*, 5376–5384.
(13) Hong, C.; Yang, S.; Ndukaife, J. C. Stand-off Trapping and Manipulation of Sub-10 Nm Objects and Biomolecules Using Opto-Thermo-Electrohydrodynamic Tweezers. *Nat. Nanotechnol.* **2020**, *15*, 908–913.
(14) Krishnan, M.; Mojarad, N.; Kukura, P.; Sandoghdar, V. Geometry-Induced Electrostatic Trapping of Nanometric Objects in a Fluid. *Nature* **2010**, *467*, 692–695.
(15) Ruggeri, F.; Zosel, F.; Mutter, N.; Różycka, M.; Wojtas, M.; Ożyhar, A.; Schuler, B.; Krishnan, M. Single-Molecule Electrometry. *Nat. Nanotechnol.* **2017**, *12*, 488–495.
(16) Zhang, S.; Xu, B.; Elsayed, M.; Nan, F.; Liang, W.; Valley, J. K.; Liu, L.; Huang, Q.; Wu, M. C.; Wheeler, A. R. Optoelectronic Tweezers: A Versatile Toolbox for Nano-/Micro-Manipulation. *Chem. Soc. Rev.* **2022**, *51*, 9203–9242.
(17) Nan, F.; Yan, Z. Sorting Metal Nanoparticles with Dynamic and Tunable Optical Driven Forces. *Nano Lett.* **2018**, *18*, 4500–4505.
(18) Kotnala, A.; Kollipara, P. S.; Li, J.; Zheng, Y. Overcoming Diffusion-Limited Trapping in Nanoaperture Tweezers Using Opto-Thermal-Induced Flow. *Nano Lett.* **2020**, *20*, 768–779.
(19) Lin, L.; Zhang, J.; Peng, X.; Wu, Z.; Coughlan, A. C. H.; Mao, Z.; Bevan, M. A.; Zheng, Y. Opto-Thermophoretic Assembly of Colloidal Matter. *Sci. Adv.* **2017**, *3*, e1700458.
(20) Svirelis, J.; Adali, Z.; Emilsson, G.; Medin, J.; Andersson, J.; Vattikunta, R.; Hulander, M.; Järlebark, J.; Kolman, K.; Olsson, O.; Sakiyama, Y.; Lim, R. Y. H.; Dahlin, A. Stable Trapping of Multiple Proteins at Physiological Conditions Using Nanoscale Chambers with Macromolecular Gates. *Nat. Commun.* **2023**, *14*, 5131.
(21) Peng, Y.; Zhou, J.; Chen, J.; Du, P.; Jin, Z.; Dai, X.; Zhong, Y.; Ji, Y.; Chen, Z.; Wang, M.; Wang, Y.; Ho, A. H.-P.; Zeng, S.; Jiang, Q.; Ma, L.; Schmidt, O. G.; Zheng, Y.; Qu, J.; Shao, Y. Optothermal Ice–Water Interface Management for Cross-Scale Enrichment and Molecular Sensing. *ACS Nano* **2025**, *19*, 39281–39291.
(22) Braun, M.; Cichos, F. Optically Controlled Thermophoretic Trapping of Single Nano-Objects. *ACS Nano* **2013**, *7*, 11200–11208.
(23) Braun, M.; Bregulla, A. P.; Günther, K.; Mertig, M.; Cichos, F. Single Molecules Trapped by Dynamic Inhomogeneous Temperature Fields. *Nano Lett.* **2015**, *15*, 5499–5505.
(24) Lin, L.; Peng, X.; Wang, M.; Scarabelli, L.; Mao, Z.; Liz-Marzán, L. M.; Becker, M. F.; Zheng, Y. Light-Directed Reversible Assembly of Plasmonic Nanoparticles Using Plasmon-Enhanced Thermophoresis. *ACS Nano* **2016**, *10*, 9659–9668.
(25) Lin, L.; Peng, X.; Mao, Z.; Wei, X.; Xie, C.; Zheng, Y. Interfacial-Entropy-Driven Thermophoretic Tweezers. *Lab. Chip* **2017**, *17*, 3061–3070.
(26) Chen, J.; Cong, H.; Loo, J.; Kang, Z.; Tang, M.; Zhang, H.; Wu, S.-Y.; Kong, S.-K.; Ho, H.-P. Thermal Gradient Induced Tweezers for the Manipulation of Particles and Cells. *Sci. Rep.* **2016**, *6*, 35814.
(27) Pang, Y.; Gordon, R. Optical Trapping of a Single Protein. *Nano Lett.* **2012**, *12*, 402–406.
(28) Kotnala, A.; Gordon, R. Quantification of High-Efficiency Trapping of Nanoparticles in a Double Nanohole Optical Tweezer. *Nano Lett.* **2014**, *14*, 853–856.
(29) Mestres, P.; Berthelot, J.; Aćimović, S. S.; Quidant, R. Unraveling the Optomechanical Nature of Plasmonic Trapping. *Light Sci. Appl.* **2016**, *5*, e16092.
(30) Crozier, K. B. Quo Vadis, Plasmonic Optical Tweezers? *Light Sci. Appl.* **2019**, *8*, 35.
(31) Jiang, Q.; Roy, P.; Claude, J.-B.; Wenger, J. Single Photon Source from a Nanoantenna-Trapped Single Quantum Dot. *Nano Lett.* **2021**, *21*, 7030–7036.

(32) Bouloumis, T. D.; Kotsifaki, D. G.; Nic Chormaic, S. Enabling Self-Induced Back-Action Trapping of Gold Nanoparticles in Metamaterial Plasmonic Tweezers. *Nano Lett.* **2023**, *23*, 4723–4731.
(33) Kotsifaki, D. G.; Truong, V. G.; Chormaic, S. N. Fano-Resonant, Asymmetric, Metamaterial-Assisted Tweezers for Single Nanoparticle Trapping. *Nano Lett.* **2020**, *20*, 3388–3395.
(34) Cohen, A. E.; Moerner, W. E. Suppressing Brownian Motion of Individual Biomolecules in Solution. *Proc. Natl. Acad. Sci.* **2006**, *103*, 4362–4365.
(35) Wang, Q.; Goldsmith, R. H.; Jiang, Y.; Bockenhauer, S. D.; Moerner, W. E. Probing Single Biomolecules in Solution Using the Anti-Brownian Electrokinetic (ABEL) Trap. *Acc. Chem. Res.* **2012**, *45*, 1955–1964.
(36) Würger, A. Thermal Non-Equilibrium Transport in Colloids. *Rep. Prog. Phys.* **2010**, *73*, 126601.
(37) Kotsifaki, D. G.; Chormaic, S. N. The Role of Temperature-Induced Effects Generated by Plasmonic Nanostructures on Particle Delivery and Manipulation: A Review. *Nanophotonics* **2022**, *11*, 2199–2218.
(38) Niether, D.; Wiegand, S. Thermophoresis of Biological and Biocompatible Compounds in Aqueous Solution. *J. Phys. Condens. Matter* **2019**, *31*, 503003.
(39) Simon, D. J.; Hartmann, F.; Thalheim, T.; Cichos, F. Thermoplasmonic Manipulation for the Study of Single Polymers and Protein Aggregates. *Macromol. Chem. Phys.* **2023**, *224*, 2300060.
(40) Braun, D.; Libchaber, A. Trapping of DNA by Thermophoretic Depletion and Convection. *Phys. Rev. Lett.* **2002**, *89*, 188103.
(41) Jiang, H.-R.; Wada, H.; Yoshinaga, N.; Sano, M. Manipulation of Colloids by a Nonequilibrium Depletion Force in a Temperature Gradient. *Phys. Rev. Lett.* **2009**, *102*, 208301.
(42) Maeda, Y. T.; Buguin, A.; Libchaber, A. Thermal Separation: Interplay between the Soret Effect and Entropic Force Gradient. *Phys. Rev. Lett.* **2011**, *107*, 038301.
(43) Maeda, Y. T.; Tlusty, T.; Libchaber, A. Effects of Long DNA Folding and Small RNA Stem–Loop in Thermophoresis. *Proc. Natl. Acad. Sci.* **2012**, *109*, 17972–17977.
(44) Simon, D. J.; Thalheim, T.; Cichos, F. Accumulation and Stretching of DNA Molecules in Temperature-Induced Concentration Gradients. *J. Phys. Chem. B* **2023**, *127*, 10861–10870.
(45) Chen, J.; Zhou, J.; Peng, Y.; Dai, X.; Tan, Y.; Zhong, Y.; Li, T.; Zou, Y.; Hu, R.; Cui, X.; Ho, H.-P.; Gao, B. Z.; Zhang, H.; Chen, Y.; Wang, M.; Zhang, X.; Qu, J.; Shao, Y. Highly-Adaptable Optothermal Nanotweezers for Trapping, Sorting, and Assembling across Diverse Nanoparticles. *Adv. Mater.* **2024**, *36*, 2309143.
(46) Fränzl, M.; Cichos, F. Hydrodynamic Manipulation of Nano-Objects by Optically Induced Thermo-Osmotic Flows. *Nat. Commun.* **2022**, *13*, 656.
(47) Jiang, Q.; Rogez, B.; Claude, J.-B.; Baffou, G.; Wenger, J. Quantifying the Role of the Surfactant and the Thermophoretic Force in Plasmonic Nano-Optical Trapping. *Nano Lett.* **2020**, *20*, 8811–8817.
(48) Anyika, T.; Hong, I.; Zhu, G.; Ndukaife, J. C. Application of Resonant Plasmonic Bowtie Nanoantennas for Optically-Assisted Diffusiophoretic Trapping of Extracellular Vesicles and Nanoparticles. *Laser Photonics Rev.* **2025**, *19*, 2400412.
(49) Zhang, F.; Oiticica, P. R. A.; Abad-Arredondo, J.; Arai, M. S.; Oliveira, O. N. Jr.; Jaque, D.; Fernandez Dominguez, A. I.; de Camargo, A. S. S.; Haro-González, P. Brownian Motion Governs the Plasmonic Enhancement of Colloidal Upconverting Nanoparticles. *Nano Lett.* **2024**, *24*, 3785–3792.
(50) Bregulla, A. P.; Würger, A.; Günther, K.; Mertig, M.; Cichos, F. Thermo-Osmotic Flow in Thin Films. *Phys. Rev. Lett.* **2016**, *116*, 188303.
(51) Wohland, T.; Maiti, S.; Machan, R. *An Introduction to Fluorescence Correlation Spectroscopy*; IOP Publishing Ltd, 2020.
(52) Haustein, E.; Schwille, P. Ultrasensitive Investigations of Biological Systems by Fluorescence Correlation Spectroscopy. *Methods* **2003**, *29*, 153–166.
(53) Jiang, Q.; Claude, J.-B.; Wenger, J. Plasmonic Nano-Optical Trap Stiffness Measurements and Design Optimization. *Nanoscale* **2021**, *13*, 4188–4194.

(54) Bar-Ziv, R.; Meller, A.; Tlusty, T.; Moses, E.; Stavans, J.; Safran, S. A. Localized Dynamic Light Scattering: Probing Single Particle Dynamics at the Nanoscale. *Phys. Rev. Lett.* **1997**, *78*, 154–157.
(55) Meller, A.; Bar-Ziv, R.; Tlusty, T.; Moses, E.; Stavans, J.; Safran, S. A. Localized Dynamic Light Scattering: A New Approach to Dynamic Measurements in Optical Microscopy. *Biophys. J.* **1998**, *74*, 1541–1548.
(56) Viana, N. B.; Freire, R. T. S.; Mesquita, O. N. Dynamic Light Scattering from an Optically Trapped Microsphere. *Phys. Rev. E* **2002**, *65*, 041921.
(57) Ito, S.; Sugiyama, T.; Toitani, N.; Katayama, G.; Miyasaka, H. Application of Fluorescence Correlation Spectroscopy to the Measurement of Local Temperature in Solutions under Optical Trapping Condition. *J. Phys. Chem. B* **2007**, *111*, 2365–2371.
(58) Jiang, Q.; Rogez, B.; Claude, J.-B.; Baffou, G.; Wenger, J. Temperature Measurement in Plasmonic Nanoapertures Used for Optical Trapping. *ACS Photonics* **2019**, *6*, 1763–1773.
(59) Jiang, Q.; Rogez, B.; Claude, J.-B.; Moreau, A.; Lumeau, J.; Baffou, G.; Wenger, J. Adhesion Layer Influence on Controlling the Local Temperature in Plasmonic Gold Nanoholes. *Nanoscale* **2020**, *12*, 2524–2531.
(60) Braun, D.; Libchaber, A. Trapping of DNA by Thermophoretic Depletion and Convection. *Phys. Rev. Lett.* **2002**, *89*, 188103.
(61) Polyanskiy, M. N. Refractiveindex.Info Database of Optical Constants. *Sci. Data* **2024**, *11*, 94.
(62) Baffou, G.; Cichos, F.; Quidant, R. Applications and Challenges of Thermoplasmonics. *Nat. Mater.* **2020**, *19*, 946–958.
(63) Xu, Z.; Song, W.; Crozier, K. B. Direct Particle Tracking Observation and Brownian Dynamics Simulations of a Single Nanoparticle Optically Trapped by a Plasmonic Nanoaperture. *ACS Photonics* **2018**, *5*, 2850–2859.
(64) Rigneault, H.; Lenne, P.-F. Fluorescence Correlation Spectroscopy on a Mirror. *JOSA B* **2003**, *20*, 2203–2214.
(65) Novotny, L.; Hecht, B. *Principles of Nano-Optics*; Cambridge University Press, 2012.
(66) Chan, J.; Popov, J. J.; Kolisnek-Kehl, S.; Leaist, D. G. Soret Coefficients for Aqueous Polyethylene Glycol Solutions and Some Tests of the Segmental Model of Polymer Thermal Diffusion. *J. Solut. Chem.* **2003**, *32*, 197–214.
(67) Braibanti, M.; Vigolo, D.; Piazza, R. Does Thermophoretic Mobility Depend on Particle Size? *Phys. Rev. Lett.* **2008**, *100*, 108303.
(68) Hänggi, P.; Talkner, P.; Borkovec, M. Reaction-Rate Theory: Fifty Years after Kramers. *Rev. Mod. Phys.* **1990**, *62*, 251–341.
(69) Wexler, D.; Gov, N.; Rasmussen, K. Ø.; Bel, G. Dynamics and Escape of Active Particles in a Harmonic Trap. *Phys. Rev. Res.* **2020**, *2*, 013003.
(70) Grebenkov, D. S. First Exit Times of Harmonically Trapped Particles: A Didactic Review. *J. Phys. Math. Theor.* **2014**, *48*, 013001.
(71) Juan, M. L.; Gordon, R.; Pang, Y.; Eftekhari, F.; Quidant, R. Self-Induced Back-Action Optical Trapping of Dielectric Nanoparticles. *Nat. Phys.* **2009**, *5*, 915–919.

## Supporting Information for

# Single Nanoparticle Dynamics in Opto-Thermal Tweezers: Resolving the Temporal Resolution of Depletion Force Trapping

Jinchao Chen,[1] Robert Talla Kontchou,[2] Saurabh Rai,[1] Guillaume Baffou,[1] Sylvain Blaize,[2] Quanbo Jiang,[2] and Jérôme Wenger[1,*]

[1] *Aix Marseille Univ, CNRS, Centrale Med, Institut Fresnel, AMUTech, 13013 Marseille, France*

[2] *Light, Nanomaterials, Nanotechnologies (L2n) Laboratory, CNRS EMR 7004, University of Technology of Troyes, 10004 Troyes, France*

* *Corresponding author: jerome.wenger@fresnel.fr*

**Contents:**

## S1. Theory of steady-state concentration gradients

The theoretical modelling of the stationary concentration distributions used to compute the data in Fig. 4 and S9 follows the approach widely used to describe thermophoresis and depletion forces.[1–6] Briefly, in presence of a temperature gradient $\nabla T$, the flux of PEG molecules contains contributions of diffusion and thermophoresis following the relation[2,3,5]

$$j_{PEG} = -D^{PEG}\nabla c_{PEG} - c_{PEG}D_T^{PEG}\nabla T \quad \text{(S1)}$$

where $c_{PEG}(r)$ is the concentration of PEG molecules, and $D^{PEG}$ and $D_T^{PEG}$ are the diffusion and thermodiffusion coefficients related to the Soret coefficient by $S_T^{PEG} = D_T^{PEG}/D^{PEG}$. Under steady-state conditions the total flux $j_{PEG}$ vanishes, and the PEG concentration is given by[2,3,5]

$$c_{PEG}(r) = c_{PEG}(\infty)\exp(-S_T^{PEG}\Delta T(r)) \quad \text{(S2)}$$

The flux of nanoparticles experiences diffusion and thermophoresis similar to Eq. (S1) for the PEG molecules. However, due to the presence of the concentration gradient of PEG molecules, depletion forces are generated on the nanoparticles, inducing a Supporting diffusiophoresis term: [1–6]

$$j_{NP} = -D^{NP}\nabla c_{NP} - c_{NP}D_T^{NP}\nabla T + c_{NP}\,u \quad \text{(S3)}$$

where $u(r) = \frac{k_B T}{3\eta}(S_T^{PEG} - 1/T)\lambda^2 c_{PEG}(r)\,\nabla T(r)$ is the diffusiophoretic velocity of the nanoparticles,[1–4] with $k_B$ the Boltzmann constant, $\eta$ the viscosity of the solution and $\lambda$ denotes here the depletion depth of PEG from the nanoparticle surface, assumed to be equivalent to the gyration radius $R_G = 4.7$nm of the PEG polymer.[5,6] At the steady state $j_{NP} = 0$, the nanoparticle concentration is

$$c_{NP}(r) = c_{NP}(\infty)\exp(-S_T^{NP}\Delta T(r) + (c_{PEG}(\infty) - c_{PEG}(r))V) \quad \text{(S4)}$$

with $V = 2\pi\lambda^2 R_{NP}$ where $R_{NP}$ denotes the radius of the nanoparticle.[1–6] Equation (S4) can be rewritten in the approximate form

$$c_{NP}(r) \approx c_{NP}(\infty)\exp(-S_{T,eff}^{NP}\Delta T(r)) \quad \text{(S5)}$$

to introduce the effective Soret coefficient $S_{T,eff}^{NP}$ of the nanoparticles given by[1,2,5]

$$S_{T,eff}^{NP} = S_T^{NP} - (S_T^{PEG} - 1/T)Vc_{PEG}(\infty) \quad \text{(S6)}$$

This effective Soret coefficient shows that in presence of PEG inducing depletion forces, the term $(S_T^{PEG} - 1/T)Vc_{PEG}(\infty)$ related to diffusiophoresis can compensate the intrinsic thermophoretic coefficient $S_T^{NP}$ of the nanoparticles. As a result, the apparent behavior of the nanoparticles can look thermophilic ($S_{T,eff}^{NP} < 0$) despite the thermophobic response ($S_T^{NP} > 0$) of the nanoparticle in absence of PEG. The resulting effect on the nanoparticle concentration distribution Eq. (S4) is a local increase of $c_{NP}$ due to the positive term $\left(c_{PEG}(\infty) - c_{PEG}(r)\right)V$ compensating the $-S_T^{NP}\Delta T(r)$ intrinsic thermophoresis response of the nanoparticles.

Note that Equation (S5) has strong similarities with a trap potential depth determined by the effective Soret coefficient $\Delta E/k_B T = -S_{T,eff}^{NP}\Delta T(r)$. Therefore, an effective thermophilic behavior ($S_{T,eff}^{NP} < 0$, particles move towards the hot region) yields a positive trap potential enabling trapping. The trap stiffness is thus proportional to the effective Soret coefficient $S_{T,eff}^{NP}$ and to the temperature increase $\Delta T$.

## S2. Numerical computations of concentration gradients

To compute the data presented in Fig. 4 and S9, the temperature profile $\Delta T(r)$ is modeled by a Voigt function following the approach in ref.[5] using the Igor Pro v7 built-in function with a parameter profile set to 1. The peak amplitude of $\Delta T(r)$ linearly increases with the IR laser intensity $I_{IR}$ (Fig. 4a). The dependence $\Delta T(0) \approx 0.5\, I_{IR}$ is set according to the temperature measurements presented in Fig. 1c, which are based on the approach described in ref.[7,8] based on the fluorescence lifetime of Alexa Fluor 647 dyes.

The Soret coefficient for the 10kDa PEG molecules is set to $S_T^{PEG} = 0.064$ $K^{-1}$ according to the values tabulated in ref.[9] and used in ref.[3,4] The PEG concentration profiles shown in Fig. 4b are computed using Eq. (S2).

The Soret coefficient for the 40 nm polystyrene nanoparticles is measured to be $S_T^{NP} = 0.075$ $K^{-1}$ from the experiments reported in Fig. S10. This value is consistent with the reported data in the literature: for instance, Piazza and coworkers found 0.07 $K^{-1}$,[10] while the data obtained from Sano et al[2] and Libchaber et al[3] give respectively 0.14 and 0.33 $K^{-1}$ for 40 nm PS nanoparticles. The nanoparticle concentration profiles shown in Fig. 4c are computed using Eq. (S4), and are used to derive the intensity gains displayed in Fig. 4d. Following the common practice in this field,[1–6] we assume that the PEG depletion depth $\lambda$ from the nanoparticle surface is approximated by the PEG gyration radius of 4.7nm.[5,6,9] To express the PEG concentration, we use the fact that a 5% PEG concentration is equivalent to a 5.3 mM polymer concentration.[5]

To compute the numerical data shown in Fig. 4e, we assume that the relative increase in the diffusion time $\tau_d/\tau_{d\,PEG\,0}^{IR\,0}$ respective to the reference in absence of PEG and IR illumination is proportional to the local concentration increase $c_{NP}/c_{NP}^{IR\,0}$ (data shown in Fig. 4d) multiplied by the viscosity increase $\eta_{PEG}/\eta_{PEG\,0}$ induced by the presence of PEG (calibrated independently in Fig. S7).

## S3. Escape time from a trap potential

The movement of a particle inside a trap potential of finite depth remains a subject of intense research.[11–13] The difficulty is that the Langevin equation cannot be solved, requiring more advanced treatments based on Fokker-Planck equation.[11,14] The mean thermal escape time from a trap potential of finite depth $\Delta E$ can be approximated by Kramers escape time[12]

$$\tau_{esc} = 2\pi \sqrt{\frac{m}{k}} \exp(\frac{\Delta E}{k_B T}) \tag{S7}$$

where $m$ is the mass of the particle, $k$ the trap stiffness and $k_B$ the Boltzmann constant. In the case of a trap potential with finite characteristic radial length $L$ and small stiffness, the mean thermal escape time can be simplified to[13]

$$\tau_{esc} \approx \frac{L^2}{2D}\left(1 + \frac{1/_2 kL^2}{k_B T}\right) \approx \tau_{d}{}_{PEG\ 0}^{IR\ 0} \ (1 + \frac{\Delta E}{k_B T}) \tag{S8}$$

where $\tau_{d}{}_{PEG\ 0}^{IR\ 0} \approx \frac{L^2}{2D}$ corresponds to the diffusion time over a circle of radius $L$ for a particle with diffusion coefficient $D$ given by Stokes-Einstein law. We assume that the FCS diffusion time $\tau_d$ observed in the case of the optothermal trap is equivalent to the mean escape time $\tau_{esc}$. For small trapping stiffness such as the ones used in this study, Equation (S8) justifies that the ratio $\tau_d/\tau_{d}{}_{PEG\ 0}^{IR\ 0}$ computed in Fig. 4e provides a way to assess the trap potential $\Delta E$.

## S4. Average number of nanoparticles from FCS analysis confirms single nanoparticle nature

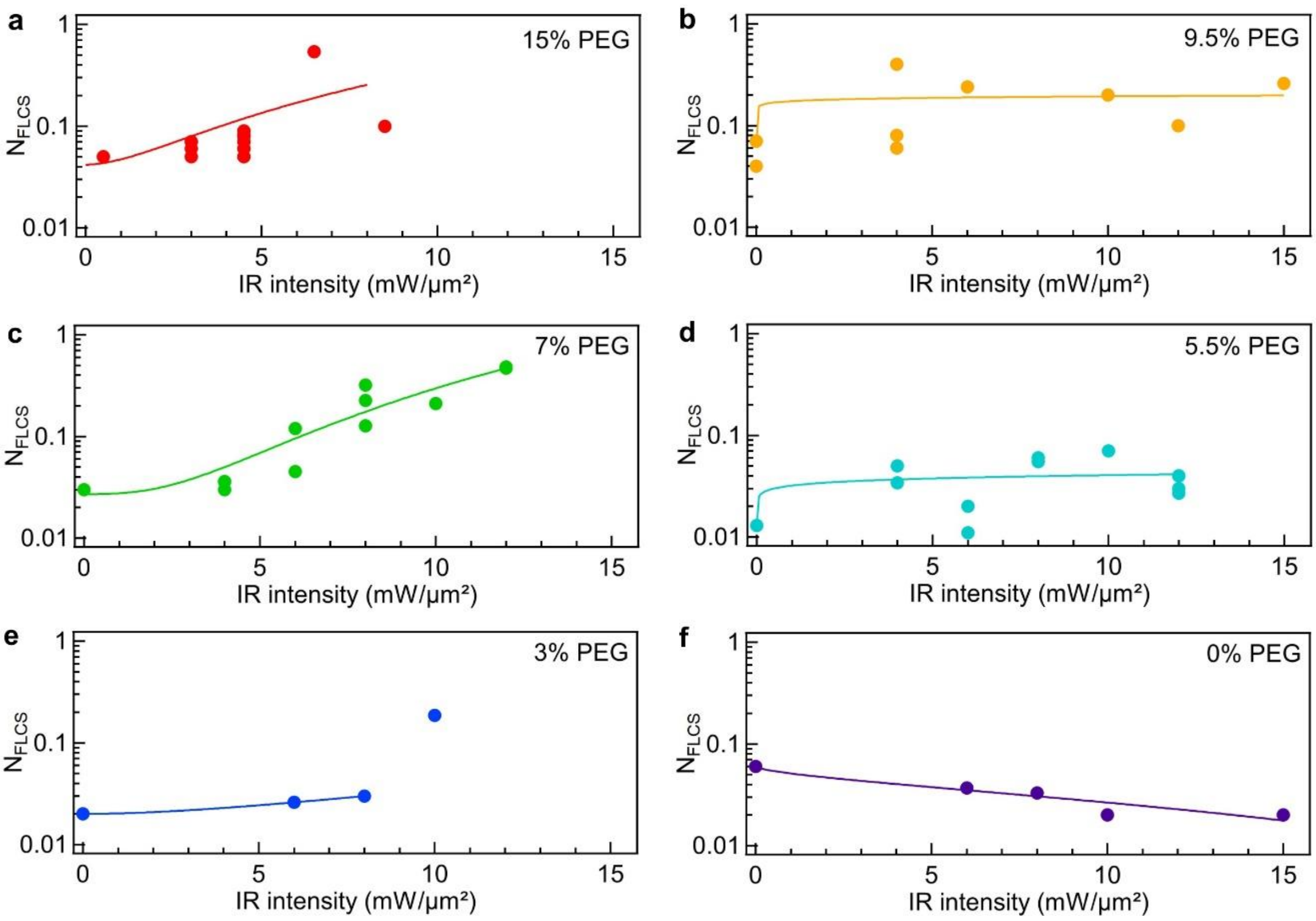

**Figure S1.** Average number of nanoparticles in the confocal detection volume determined by the FCS analysis as a function of the IR intensity and for different PEG concentrations (a-f). In all cases, the average number of nanoparticles remains below 1, indicating the single-nanoparticle nature of our results.

## S5. Supplementary fluorescence burst duration histograms

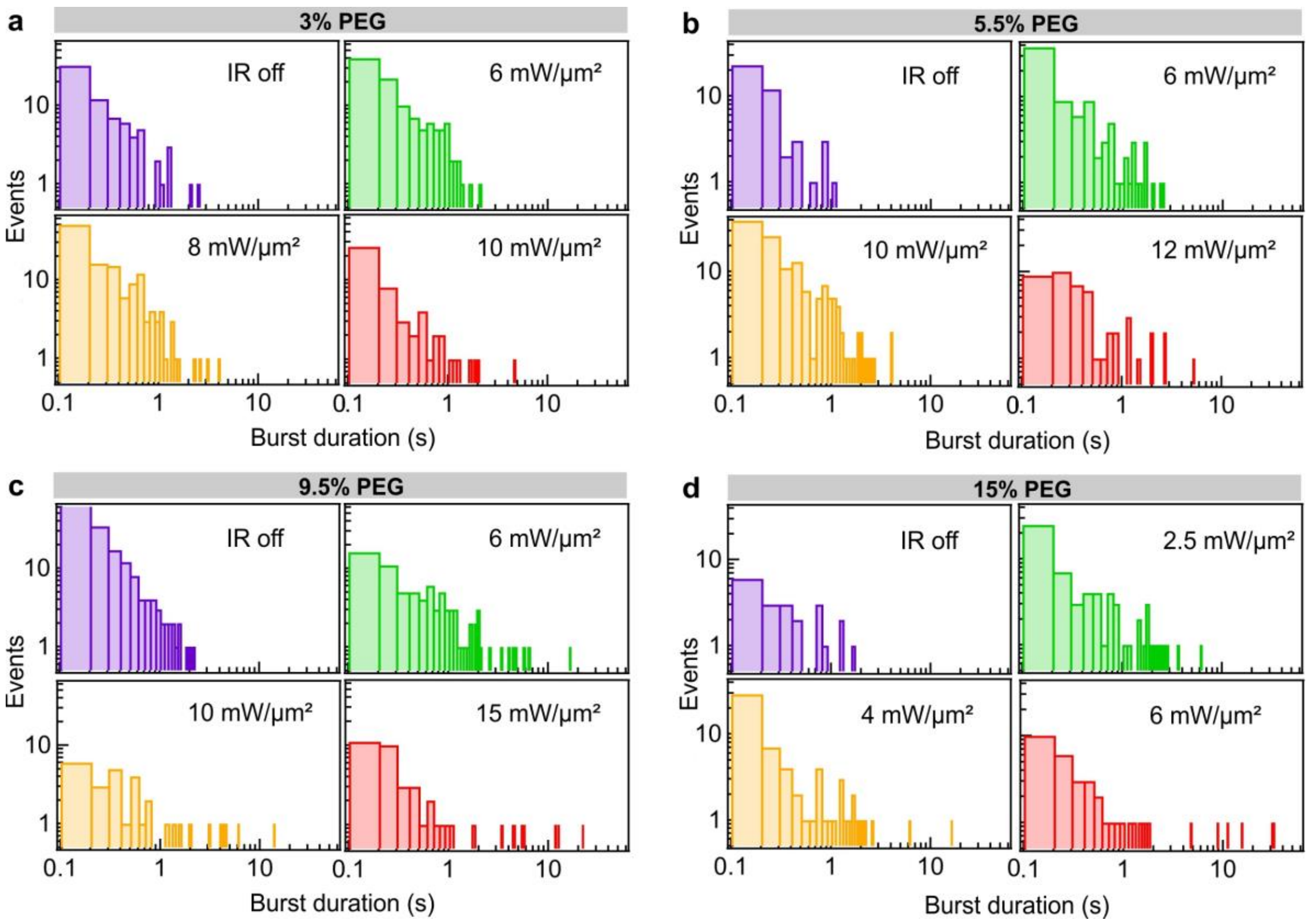

**Figure S2.** Histograms of fluorescence burst durations recorded for 3% (a), 5.5% (b), 9.5% (c) and 15% (d) PEG concentrations at different IR intensities. The conditions and analysis are similar to the ones used for Fig. 2a,b.

## S6. Supplementary complementary cumulative distribution functions

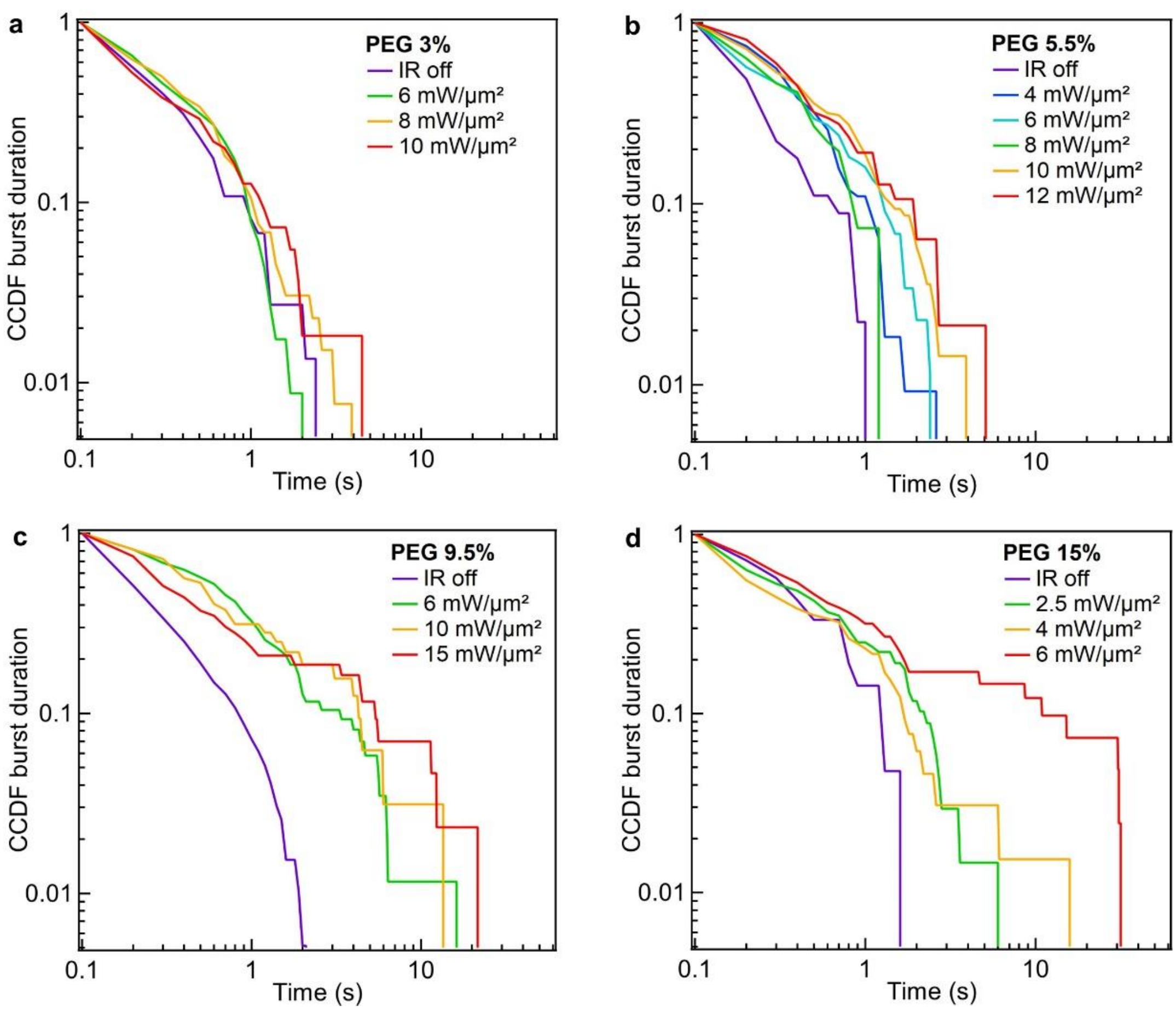

**Figure S3.** Complementary cumulative distribution functions (CCDFs) computed from the histograms in Fig. S2 for different PEG concentrations (a-d) and IR intensities.

## S7. Supplementary FCS data for different PEG concentrations

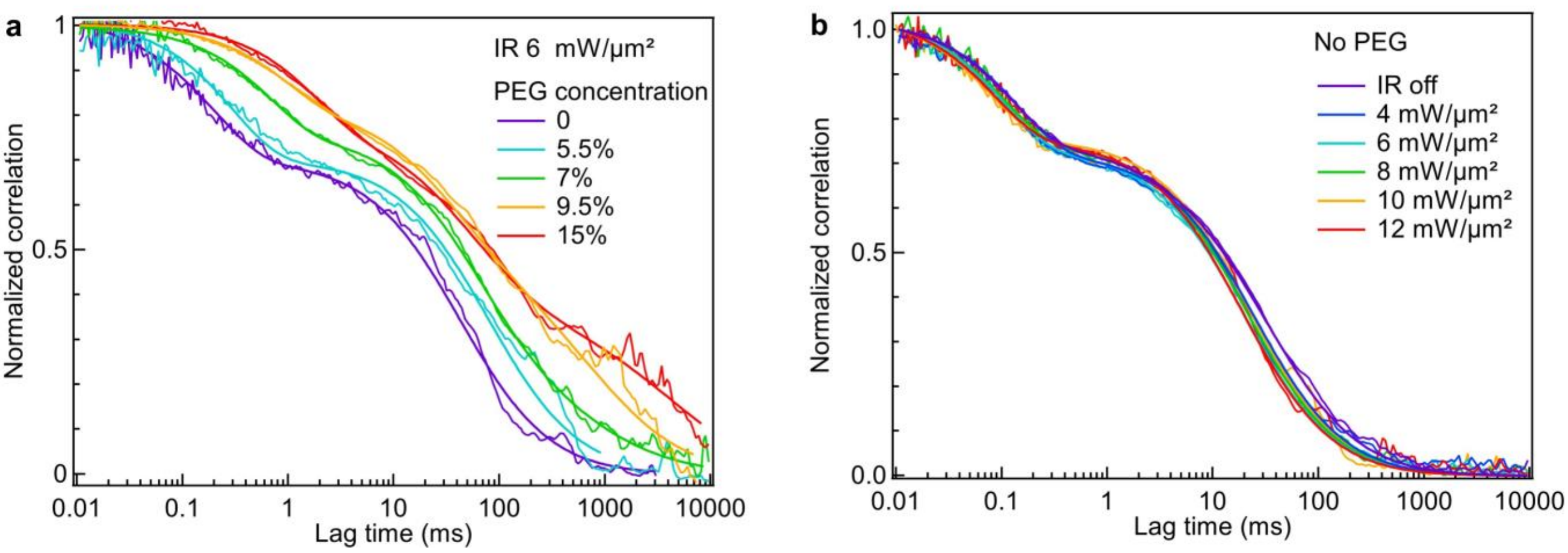

**Figure S4.** (a) Normalized FCS data (thin lines) and numerical fits (thick curves) for increasing PEG concentrations at 6 mW/µm² IR intensity. A clear increase in the diffusion time is observed for higher PEG concentrations, which cannot only be explained by a higher dynamic viscosity due to the presence of PEG (see Fig. S7 & S8 below). (b) Reference FCS data in absence of PEG for increasing IR intensities. Due to the occurrence of thermophoresis and a higher local temperature giving rise to a reduced viscosity of water, the diffusion time goes down when the IR intensity increases (as seen on Fig. 3b).

## S8. Interference fringes in the vicinity of the gold mirror

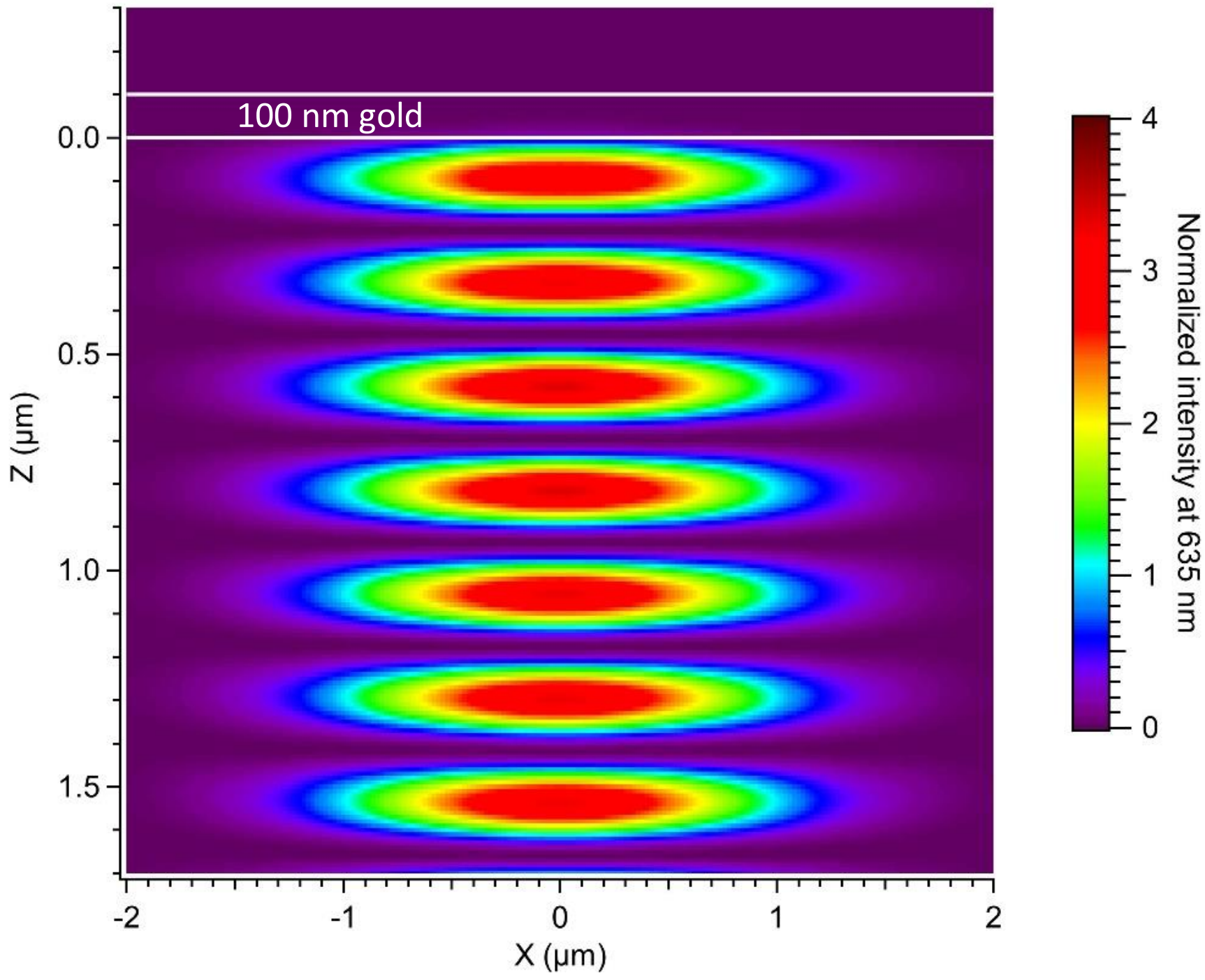

**Figure S5.** Numerical simulations of the electric field intensity $|E|^2$ distribution for the 635 nm laser impinging on the 100 nm gold mirror. The intensity has been normalized so that in the absence of the mirror the intensity in the center of the beam amounts to 1. The beam waist has been set to 1.2 µm in accordance to the experiment in Fig. S6. Interference fringes are clearly visible due to the coherent superposition of the incoming and reflected beams.

## S9. Calibration of the confocal detection volume in presence of the diaphragm

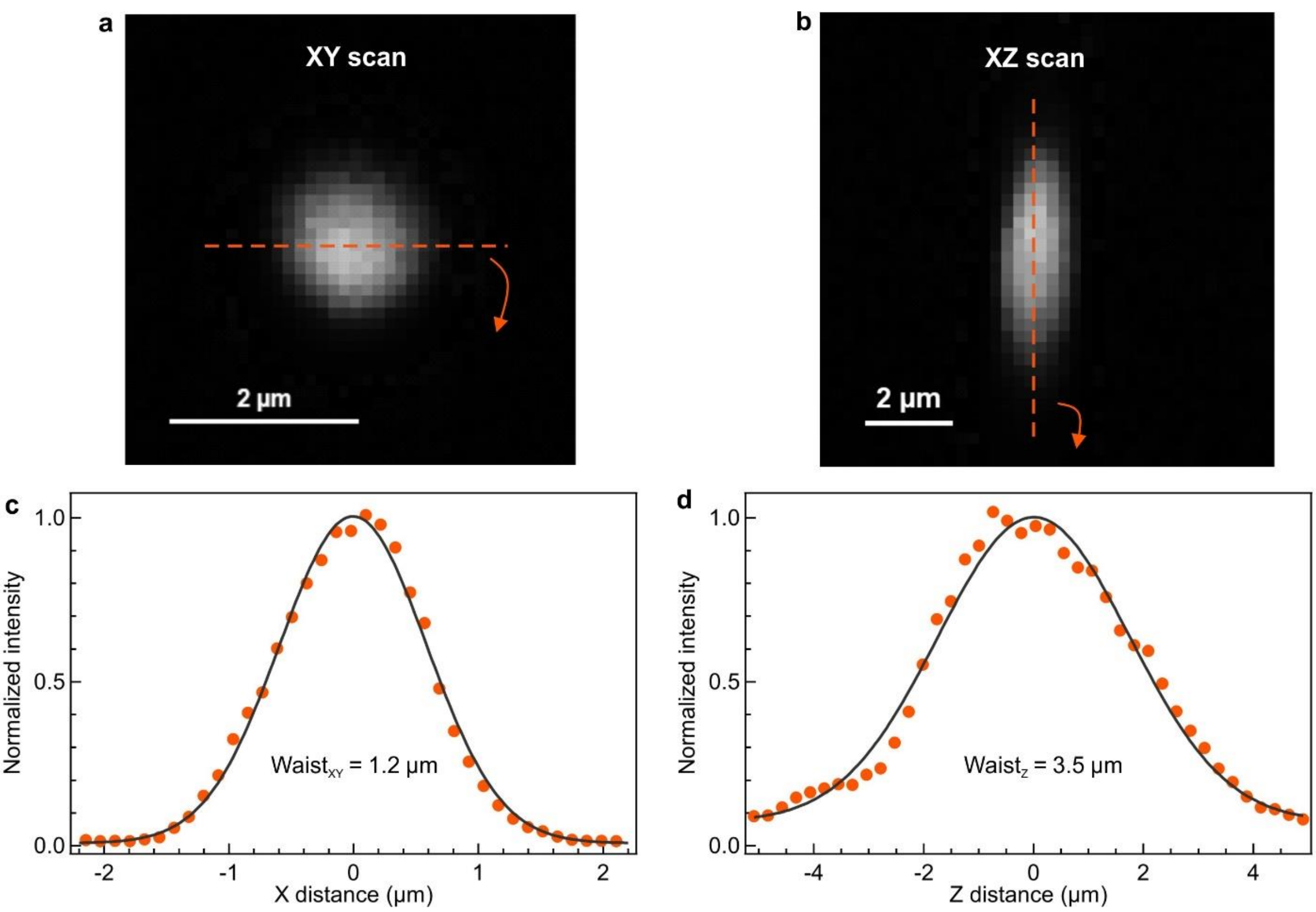

**Figure S6.** Calibration of the fluorescence microscope point spread function (PSF) with a diaphragm of 1.5 mm diameter limiting the size of the 635 nm fluorescence excitation laser beam. (a) Confocal transversal XY scanning fluorescence image of a single nanohole of 300 nm diameter milled into a 100 nm thick opaque gold film filled with Alexa Fluor 647 molecules. (b) Same as (a) for longitudinal scanning along XZ plane. (c,d) Cross-cuts along the dashed orange lines in (a,b) together with numerical fits using a Gaussian function. The waists (radius at $1/e^2$ intensity) are indicated on each graph.

## S10. Calibration of the viscosity increase in presence of PEG

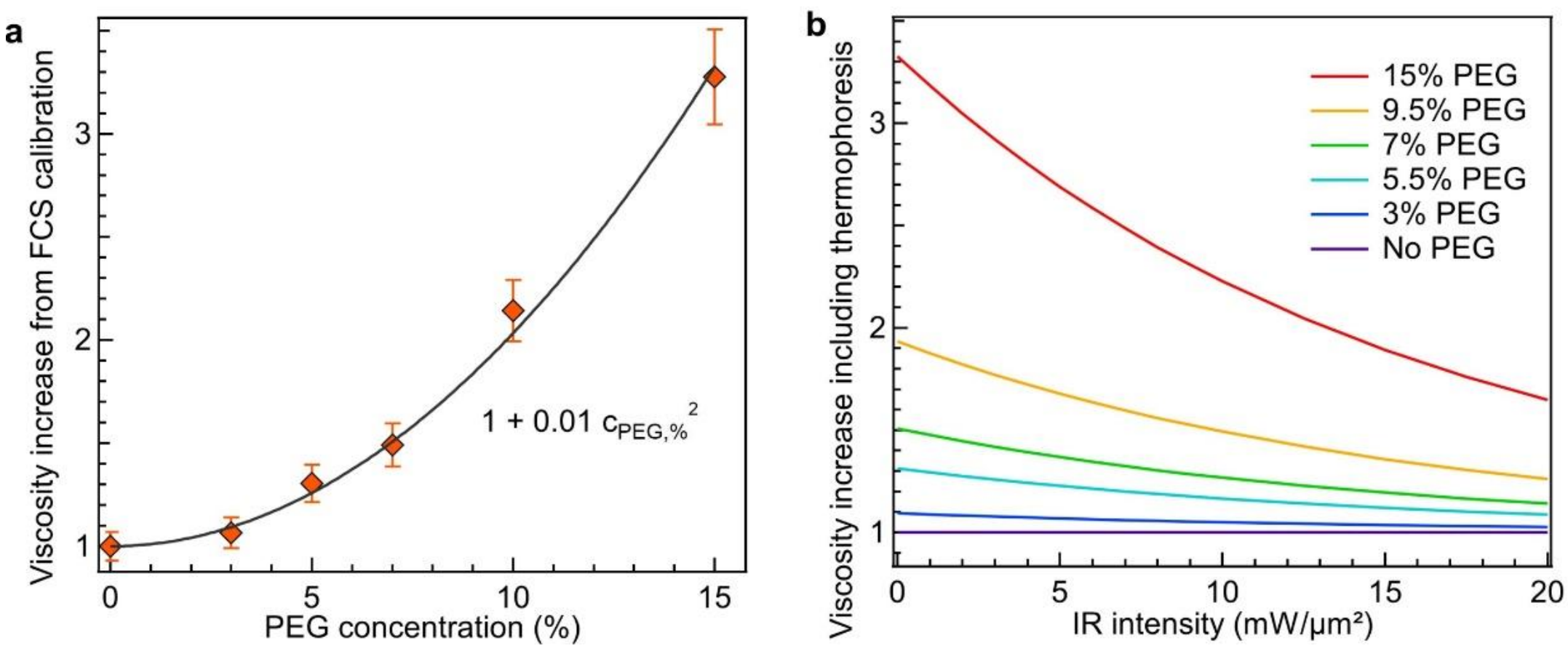

**Figure S7.** (a) Relative increase of the dynamic viscosity (determining the FCS diffusion time from the Stokes Einstein relationship) as a function of the PEG concentration. Markers are experimental data obtained from FCS calibration using CF640R fluorescent molecules on a confocal microscope. The black curve is a numerical fit. (b) Evolution of the dynamic viscosity increase as a function of the IR intensity. For these computations, we use the calibration in (a) and compensate for the lower PEG concentration induced by thermophoresis, as computed in Fig. 4b using $S_T^{PEG}$= 0.064 $K^{-1}$. These values are used to compute the evolution of the FCS diffusion time as a function of the IR intensity shown in Fig. 4e.

## S11. Influence of the viscosity on the residence time increase

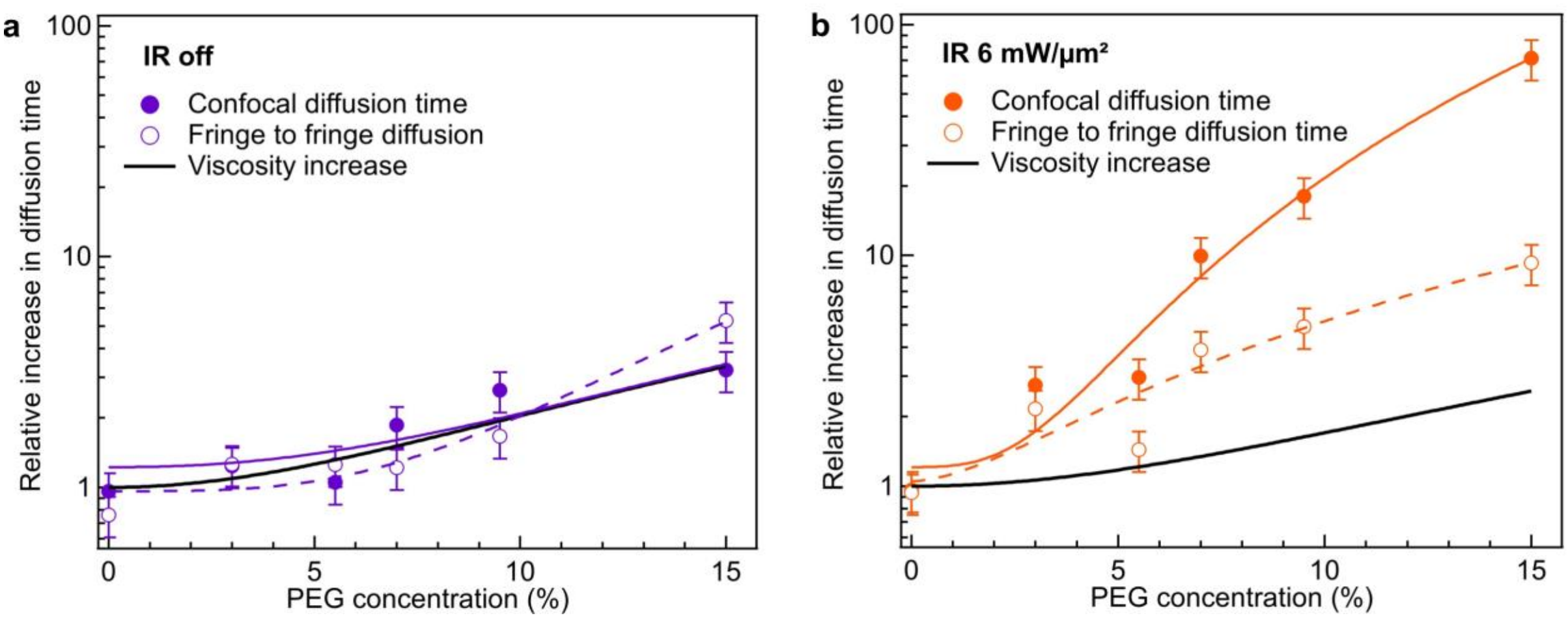

**Figure S8.** Comparison between the relative change of the FCS diffusion time $\tau_D$ (filled markers, solid purple/orange curves), the fringe-to-fringe diffusion time $\tau_f$ (empty markers, dashed purple/orange curves) and the relative increase in the dynamic viscosity (thick black line). The confocal diffusion time corresponds to the long FCS component with times in the 10s to 1000s of milliseconds. The physical picture is the time needed for the nanoparticle to escape from the 25 $\mu m^3$ confocal detection volume (calibrated in Fig. S10). The so-called fringe-to-fringe diffusion time corresponds to the short FCS component with sub-millisecond times. This is attributed to the time needed for the nanoparticle to jump from one bright interference fringe to the next one, as the presence of the gold mirror introduces an interference pattern on the 635 nm excitation profile (see Fig. S5). When the IR laser is off and the temperature is spatially uniform (a), the relative changes in both the FCS short and long diffusion times can be well accounted for by the higher dynamic viscosity related to the higher PEG concentration. When the IR intensity is 6 mW/µm² and thermal depletion forces are introduced (b), then the sole contribution of the dynamic viscosity increase due to the PEG presence can no longer explain the observed FCS times.

## S12. Simulations of the peak nanoparticle concentration gain and localization width

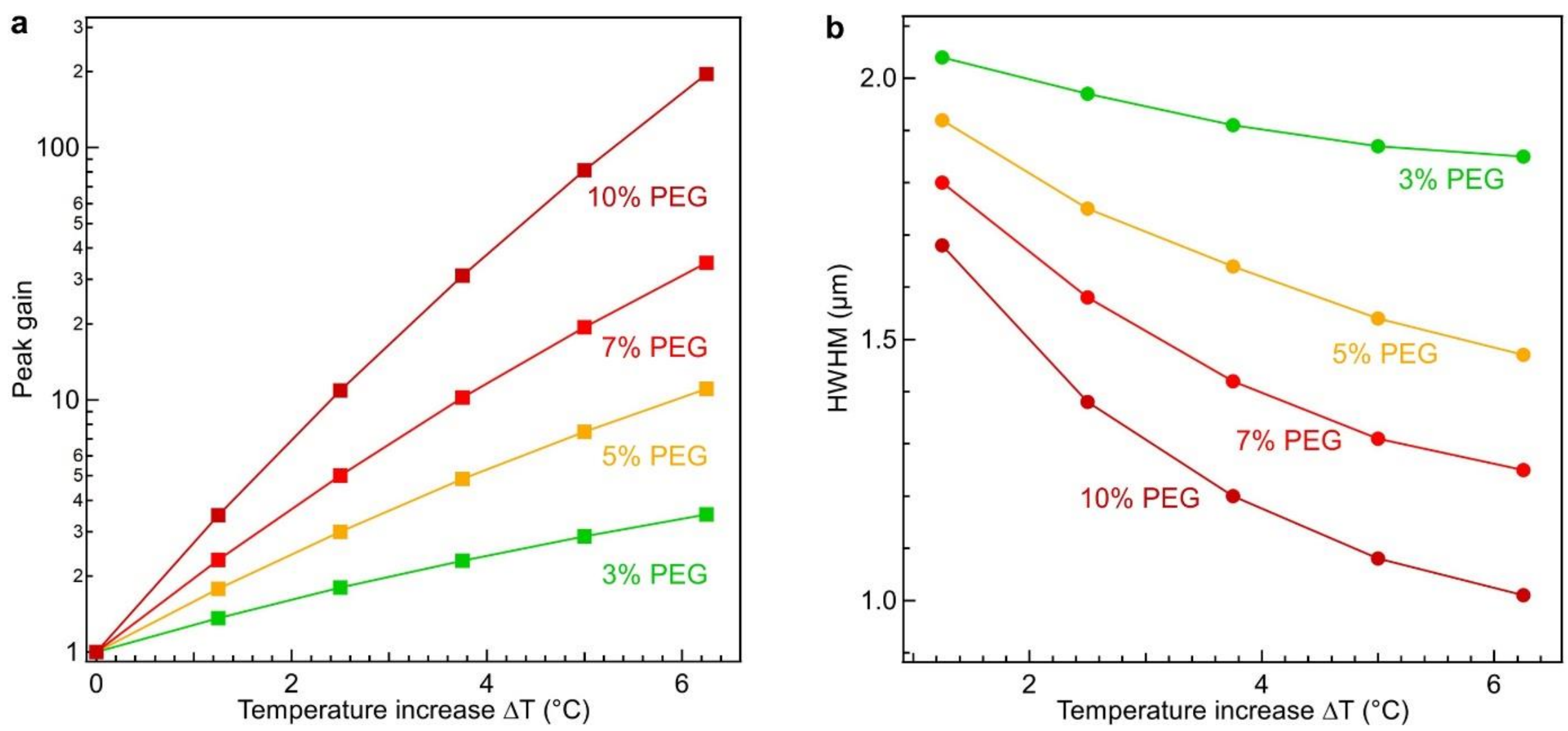

**Figure S9.** (a) Computations of the peak concentration gain for the 40 nm nanoparticles as a function of the temperature increase for different PEG concentrations. The data is extracted from simulations similar to the ones displayed in Fig. 4c. (b) Half width at half maximum (HWHM) of the nanoparticle spatial distribution (Fig. 4c) as a function of the temperature increase for different PEG concentrations. The use of higher temperature gains and higher PEG concentrations lead to a more pronounced localization of the nanoparticles respective to the IR laser hot spot.

## S13. Ensemble-averaged intensity data show accumulation

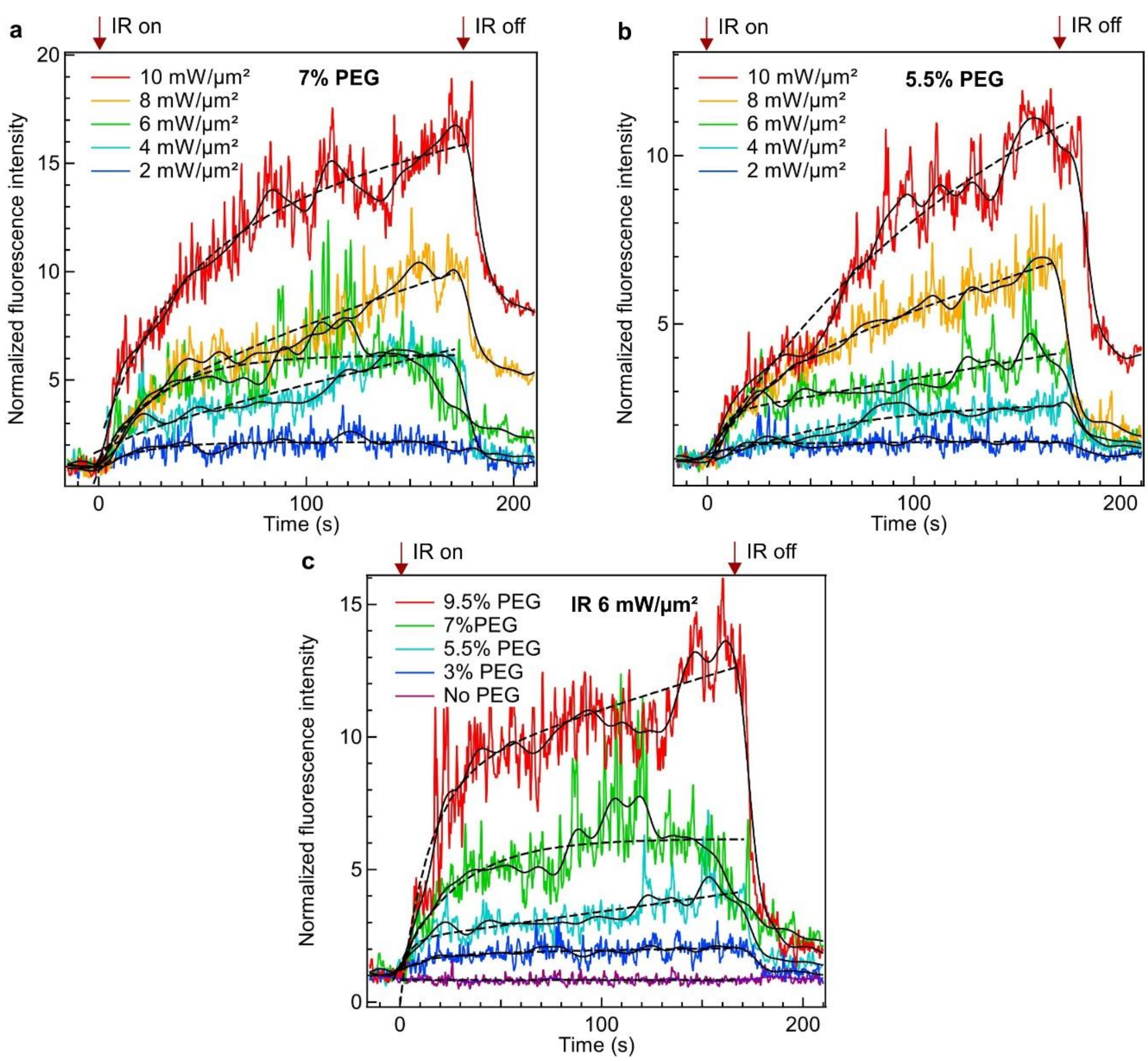

**Figure S10.** Fluorescence intensity time traces in presence of 7% PEG (a) and 5.5% PEG (b) as the IR laser is turned on at t=0 with different IR intensities, and switched off at t=180s. These experiments were recorded at a 160x higher nanoparticle concentration, to study the conditions where nanoparticles accumulate in the hot spot region. The fluorescence intensity has been normalized to the reference value when the IR laser is off (no heating). The experimental data shown in Fig. 4d correspond to the average fluorescence intensity gain between 160 and 180 s after the IR laser is switched on. (c) Fluorescence intensity time traces at an IR intensity of 6 mW/µm² for different PEG concentrations. Black solid thin lines represent the time traces after a 0.1 Hz Hanning low-pass filter has been applied to the data. Black thick dashed curves are numerical fits using a double exponential function.

## S14. Data in absence of PEG reveal the thermophobic response of the nanoparticles

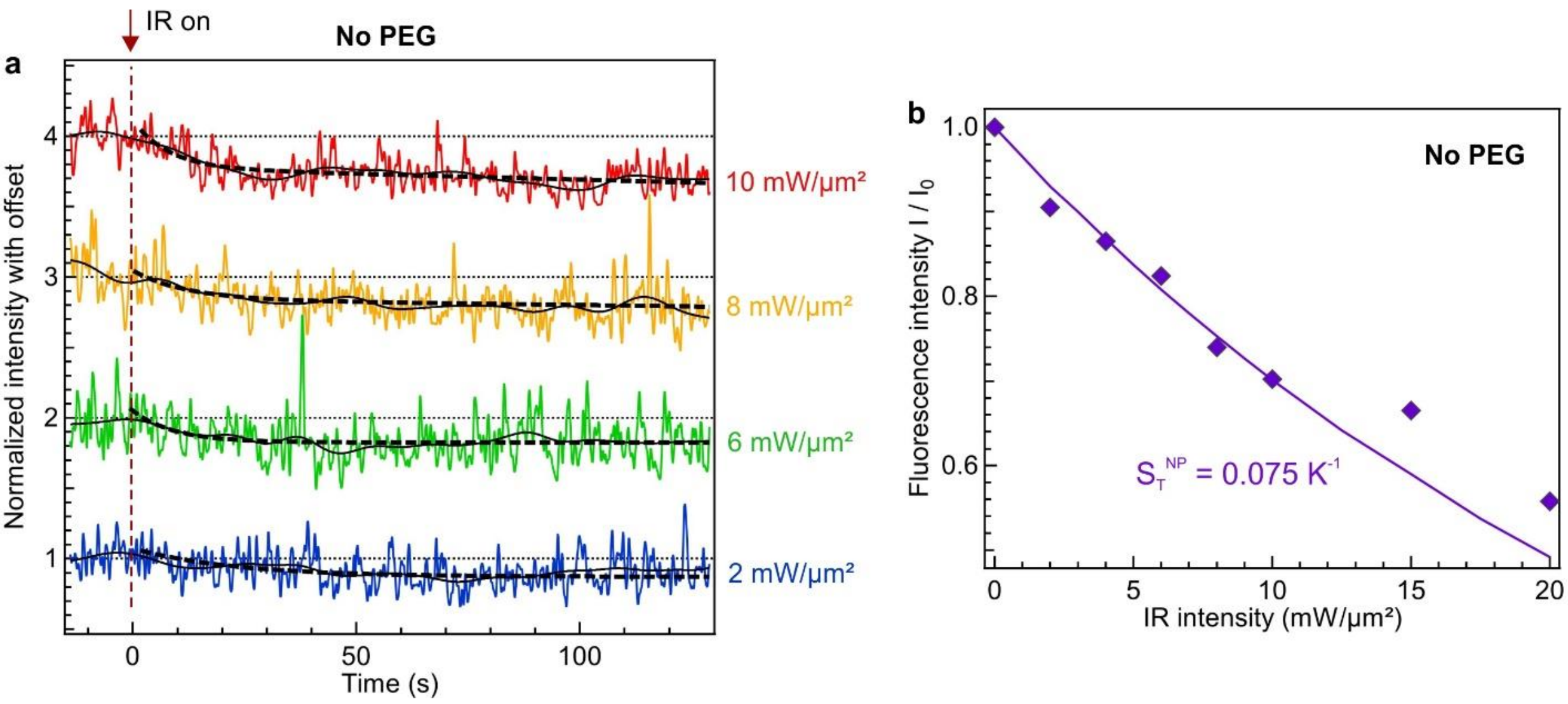

**Figure S11.** (a) Fluorescence intensity time traces in absence of PEG as the IR laser is turned on at t=0 with different IR intensities. These experiments were recorded in the similar conditions as those of Fig. S10, using a 160x higher nanoparticle concentration than the experiments used in the main article. The fluorescence intensity has been normalized to the reference value when the IR laser is off (no heating) and shifted vertically for clarity. Black solid thin lines represent the time traces after a 0.1 Hz Hanning low-pass filter has been applied to the data. Black thick dashed curves are numerical fits using a double exponential function. (b) Fluorescence intensity drop as a function of the IR intensity in the absence of PEG extracted from the time traces in (a) when the steady-state is reached (markers). Numerical simulations using the model Eq. (S2) (solid curve) are used to estimate the Soret coefficient of the 40 nm polystyrene nanoparticles to $S_T^{NP}$ = 0.075 $K^{-1}$.

## S15. Simulations of FCS diffusion times and comparison to experimental data

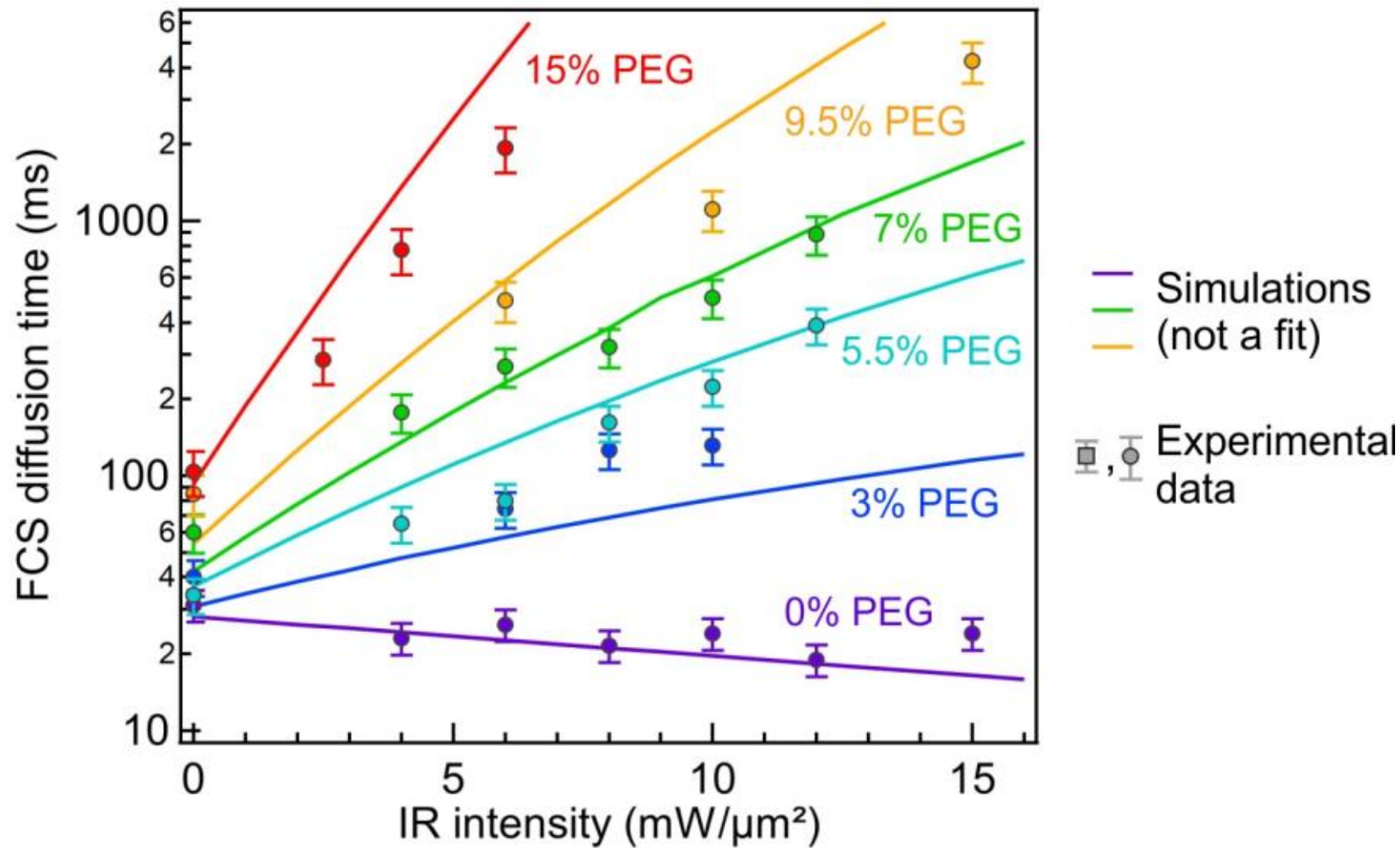

**Figure S12.** Numerical predictions (lines) and experimental data (markers, same as in Fig. 3b) of the FCS diffusion time as a function of the IR intensity for different PEG concentrations. The calculations include the calibration of the increased viscosity to the PEG local concentration (Fig. S7). Here we assume that the relative increase in the diffusion time $\tau_d/\tau_{d\,PEG\,0}^{\,IR\,0}$ respective to the reference in absence of PEG and IR illumination is proportional to the local concentration increase $c_{NP}/c_{NP}^{IR\,0}$ (data shown in Fig. 4d) multiplied by the viscosity increase $\eta_{PEG}/\eta_{PEG\,0}$ induced by the presence of PEG (calibrated independently in Fig. S7).

## S16. Experiments for 100 nm nanoparticles

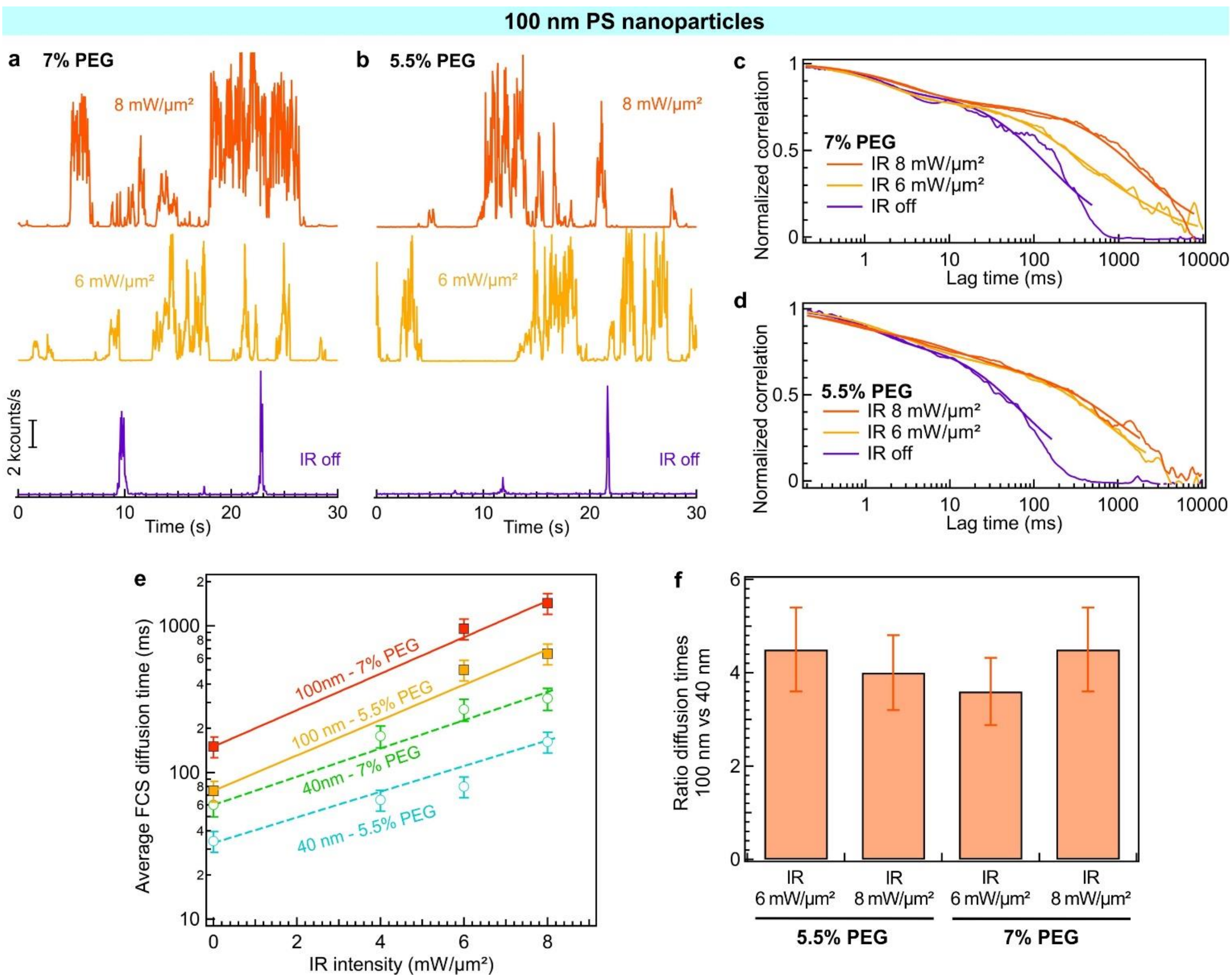

**Figure S13.** The experiments have been reproduced on 100 nm nanoparticles (ThermoFisher T7279). (a,b) Fluorescence intensity time traces for different IR intensities and PEG concentrations. The binning time is 30 ms. (c,d) Normalized FCS data (thin lines) and numerical fits (thick curves) for 7% and 5.5% PEG concentrations and different IR intensities. (e) Average diffusion time $\tau_D$ computed from the FCS fits as a function of the IR intensity for different PEG concentrations. The data for the 40 nm nanoparticles is reproduced from Fig. 3b for comparison. (f) Ratio $\tau_{d,100nm}/\tau_{d,40nm}$ of the diffusion times found for the 100 nm nanoparticles by the 40 nm result, recorded at the same IR intensity and PEG concentration. This ratio reveals the trap stiffness increase with the nanoparticle diameter. From Eq. (S6), it is expected that the effective Soret coefficient $S_{T,eff}^{NP}$ scales linearly with the nanoparticle diameter (via the inherent Soret coefficient $S_T^{NP}$ [10] and the volume exclusion $V$). Changing from 40 to 100 nm diameter thus increases $S_{T,eff}^{NP}$ by 100/40=2.5×. The larger increase in the diffusion times around 4× tends to indicate that the observed residency time scales as the power 1.5 of $S_{T,eff}^{NP}$ in our conditions (as $(2.5)^{1.5} \sim 4$).

## S17. Extension to other configurations

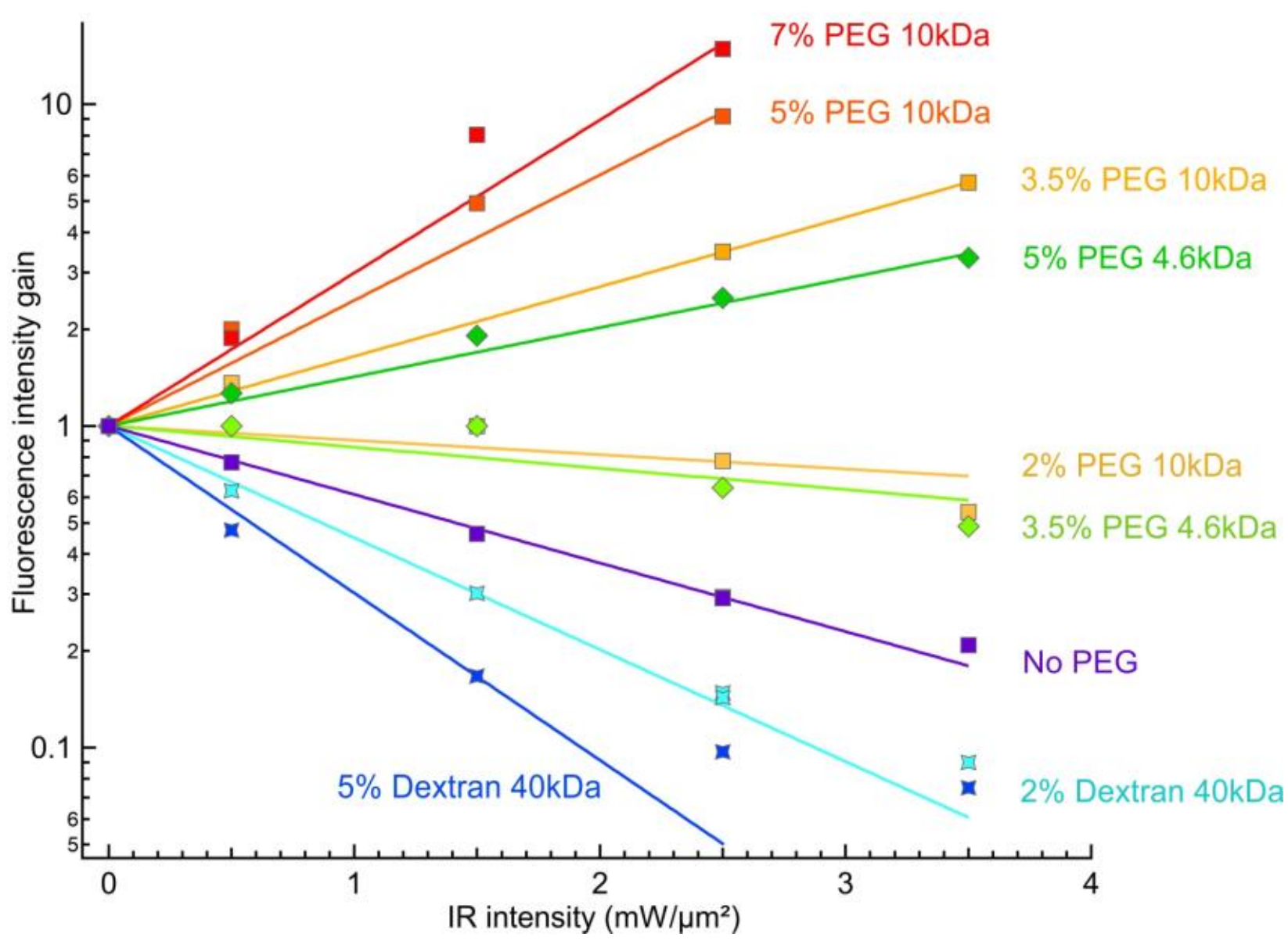

**Figure S14.** Fluorescence intensity gains deduced from experiments similar to the ones in Fig. S10 & S11 as a function of the IR intensity for different crowding agents. The measurements are performed on 28 nm PS nanoparticles at 50 nM concentration. The lines are numerical fits to the data. We observe that 4.6 kDa PEG yields qualitatively similar results to 10 kDa PEG, its lower molecular weight results in a weaker response. However, 40 kDa Dextran does not support optothermal trapping, despite its frequent use as a crowding agent.

## S18. Control for negligible photobleaching

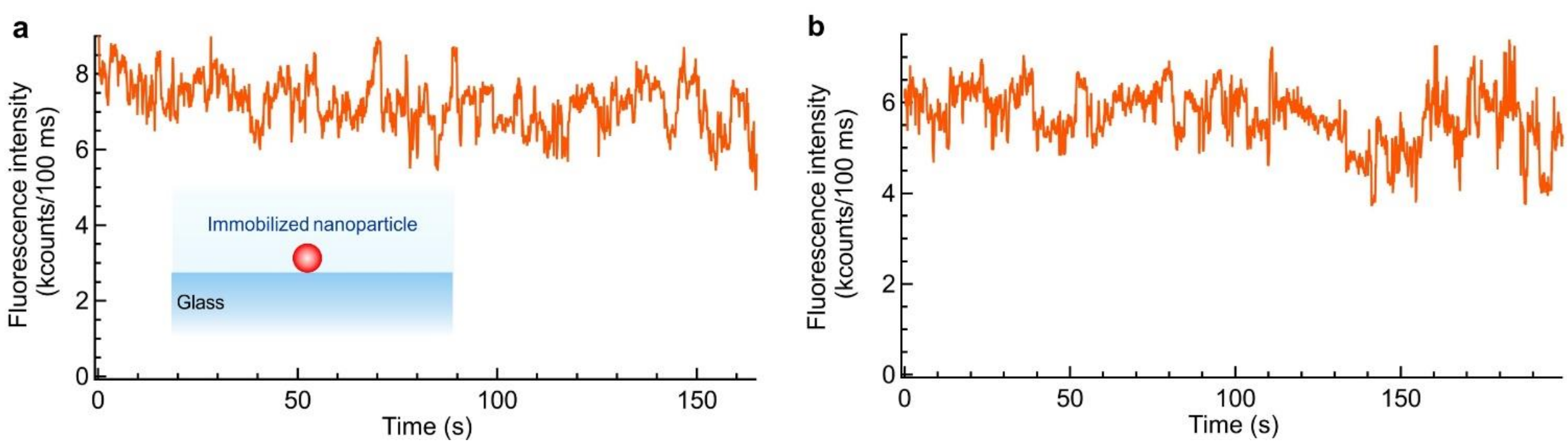

**Figure S15.** (a,b) Fluorescence intensity time traces recorded on two individual 40 nm nanoparticles immobilized on a glass coverslip in conditions similar to the ones used for optothermal trapping (Fig. 1d-f). The intensity was found stable for durations over 200 s, ruling out the hypothesis that our trapping observations are limited by photobleaching. The temporal fluctuations on the order of a few seconds are not a stability issue from our microscope, but are related to the fluorescence photodynamics of the dyes embedded into the polystyrene nanoparticle.

## S19. Control of FCS fit results

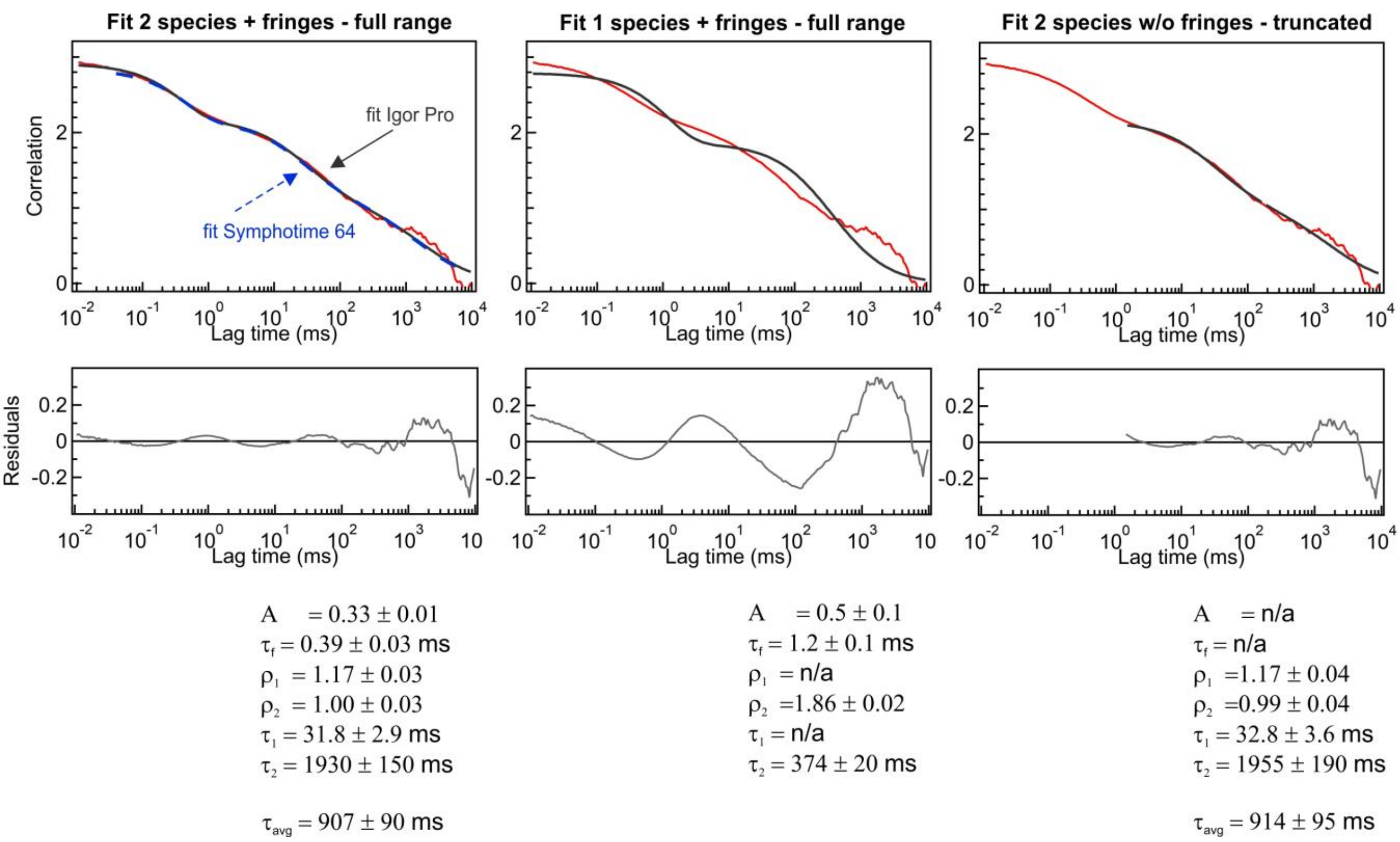

**Figure S16.** Fit results (top line) and residuals (bottom graph) for different models: 2 species with fringes term (used in the main text), single species with fringes term and truncated temporal range using 2 species only. The FCS data correspond to 7% PEG 10 kDa at 12 mW/µm² IR intensity. The fit results obtained using Symphotime 64 software (dashed blue line, top left graph) are comparable with an independent fitting using Igor Pro software (black trace).

**Supporting references:**

(1) Würger, A. Thermal Non-Equilibrium Transport in Colloids. *Rep. Prog. Phys.* **2010**, *73* (12), 126601. https://doi.org/10.1088/0034-4885/73/12/126601.
(2) Jiang, H.-R.; Wada, H.; Yoshinaga, N.; Sano, M. Manipulation of Colloids by a Nonequilibrium Depletion Force in a Temperature Gradient. *Phys. Rev. Lett.* **2009**, *102* (20), 208301. https://doi.org/10.1103/PhysRevLett.102.208301.
(3) Maeda, Y. T.; Buguin, A.; Libchaber, A. Thermal Separation: Interplay between the Soret Effect and Entropic Force Gradient. *Phys. Rev. Lett.* **2011**, *107* (3), 038301. https://doi.org/10.1103/PhysRevLett.107.038301.
(4) Maeda, Y. T.; Tlusty, T.; Libchaber, A. Effects of Long DNA Folding and Small RNA Stem–Loop in Thermophoresis. *Proc. Natl. Acad. Sci.* **2012**, *109* (44), 17972–17977. https://doi.org/10.1073/pnas.1215764109.
(5) Simon, D. J.; Thalheim, T.; Cichos, F. Accumulation and Stretching of DNA Molecules in Temperature-Induced Concentration Gradients. *J. Phys. Chem. B* **2023**, *127* (50), 10861–10870. https://doi.org/10.1021/acs.jpcb.3c06405.
(6) Chen, J.; Zhou, J.; Peng, Y.; Dai, X.; Tan, Y.; Zhong, Y.; Li, T.; Zou, Y.; Hu, R.; Cui, X.; Ho, H.-P.; Gao, B. Z.; Zhang, H.; Chen, Y.; Wang, M.; Zhang, X.; Qu, J.; Shao, Y. Highly-Adaptable Optothermal Nanotweezers for Trapping, Sorting, and Assembling across Diverse Nanoparticles. *Adv. Mater.* **2024**, *36* (9), 2309143. https://doi.org/10.1002/adma.202309143.
(7) Jiang, Q.; Rogez, B.; Claude, J.-B.; Baffou, G.; Wenger, J. Temperature Measurement in Plasmonic Nanoapertures Used for Optical Trapping. *ACS Photonics* **2019**, *6* (7), 1763–1773. https://doi.org/10.1021/acsphotonics.9b00519.
(8) Jiang, Q.; Rogez, B.; Claude, J.-B.; Moreau, A.; Lumeau, J.; Baffou, G.; Wenger, J. Adhesion Layer Influence on Controlling the Local Temperature in Plasmonic Gold Nanoholes. *Nanoscale* **2020**, *12* (4), 2524–2531. https://doi.org/10.1039/C9NR08113E.
(9) Chan, J.; Popov, J. J.; Kolisnek-Kehl, S.; Leaist, D. G. Soret Coefficients for Aqueous Polyethylene Glycol Solutions and Some Tests of the Segmental Model of Polymer Thermal Diffusion. *J. Solut. Chem.* **2003**, *32* (3), 197–214. https://doi.org/10.1023/A:1022925216642.
(10) Braibanti, M.; Vigolo, D.; Piazza, R. Does Thermophoretic Mobility Depend on Particle Size? *Phys. Rev. Lett.* **2008**, *100* (10), 108303. https://doi.org/10.1103/PhysRevLett.100.108303.
(11) Hänggi, P.; Talkner, P.; Borkovec, M. Reaction-Rate Theory: Fifty Years after Kramers. *Rev. Mod. Phys.* **1990**, *62* (2), 251–341. https://doi.org/10.1103/RevModPhys.62.251.
(12) Wexler, D.; Gov, N.; Rasmussen, K. Ø.; Bel, G. Dynamics and Escape of Active Particles in a Harmonic Trap. *Phys. Rev. Res.* **2020**, *2* (1), 013003. https://doi.org/10.1103/PhysRevResearch.2.013003.
(13) Grebenkov, D. S. First Exit Times of Harmonically Trapped Particles: A Didactic Review. *J. Phys. Math. Theor.* **2014**, *48* (1), 013001. https://doi.org/10.1088/1751-8113/48/1/013001.
(14) Novotny, L.; Hecht, B. *Principles of Nano-Optics*; Cambridge University Press, 2012.